世界互联网大会
World Internet
Conference

# Advancing a Global Framework for AI Safety and Governance for the Well-being of Humanity

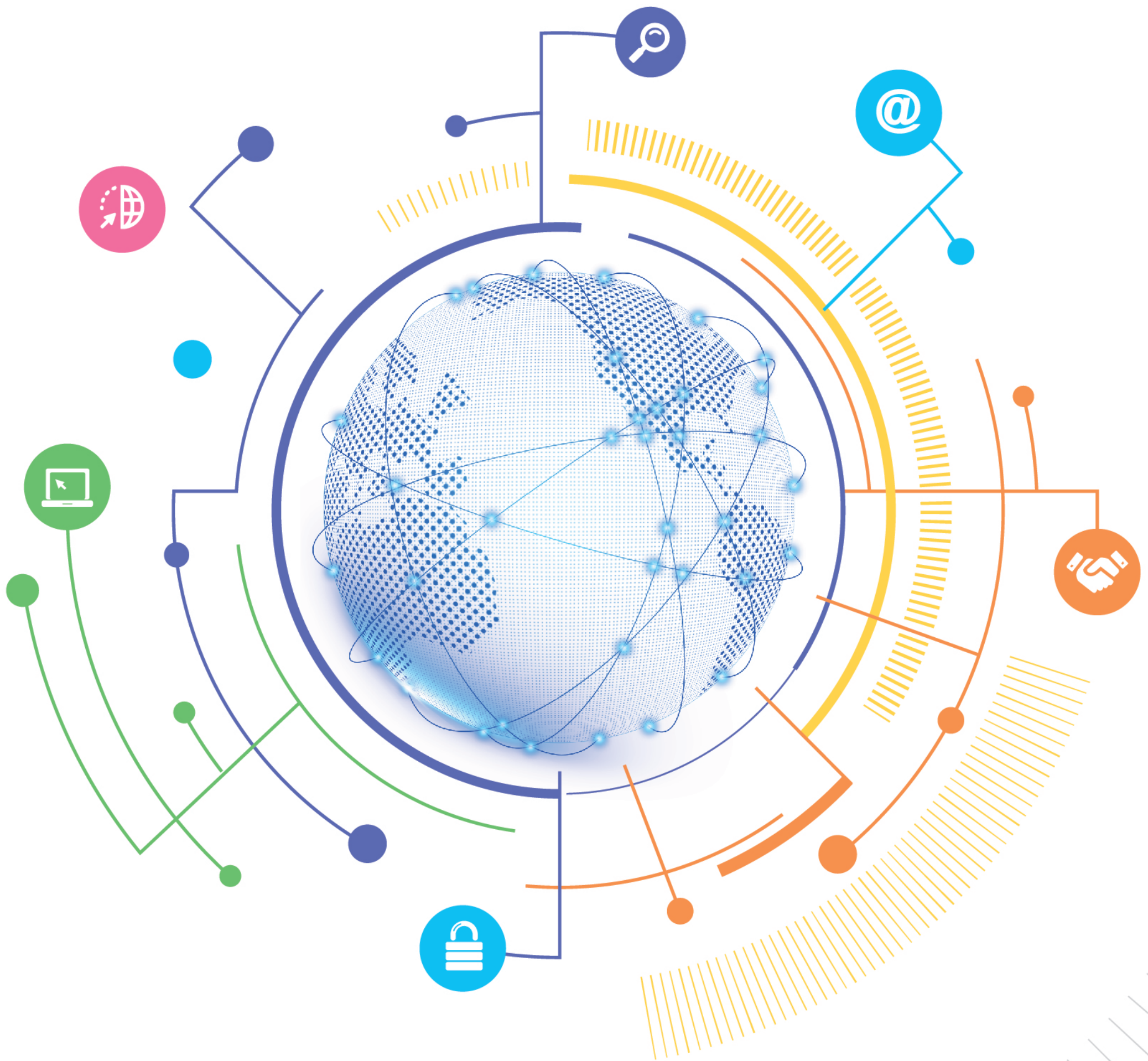

AI Safety and Governance Program of
World Internet Conference Specialized Committee on Artificial Intelligence
November 2025

## CONTRIBUTORS AND MEMBERS OF AI SAFETY AND GOVERNANCE PROGRAM OF WORLD INTERNET CONFERENCE SPECIALIZED COMMITTEE ON AI

## CO-LEADS

Zeng Yi — Chinese Academy of Sciences; Beijing Institute of AI Safety and Governance (Beijing-AISI)

Seán Ó hÉigeartaigh — University of Cambridge

## EXPERTS AND CONTRIBUTORS

Including WIC members and invited experts of the Program,
as well as contributors who provided contents or comments to the report,
listed in alphabetical order by the first letter of the organization name.

**Alibaba Cloud**
Wang Xingguang, Zhang Rong

**Arab ICT Organization (AICTO)**
Nada Laabidi

**Beijing Academy of Artificial Intelligence**
Bu Yuyan

**Beijing University of Posts and Telecommunications**
Yang Zhongliang

**Beijing Yinxiang Biji Technology Co., Ltd.**
Qiao Qian

**CCTV International Network Co., Ltd**
Cheng Ming

**China Academy of Information and Communications Technology (CAICT)**
Guo Sumin, Hu Naying

**China Branch of BRICS Institute of Future Networks**
Fan Wei, Zhou Yuan

**China Internet Network Information Center (CNNIC)**
Chen Jingjing

**China Mobile Think Tank (CMTT)**
Lin Lin, Wu Shuyan

**China Telecommunications Corporation**
Liu Weichen, Wang Feng

**China United Network Communications Group Co., Ltd.**
Zhou Kai

**China University of Political Science and Law**
Zhang Linghan

**Research and Promotion Center for Artificial Intelligence, Chinese Academy of Social Sciences**
Duan Weiwen

**Cloud Security Alliance (CSA)**
John Yeoh

**Cyber Security Association of China (CSAC)**
Wang Jianbing, Xia Wenhui

**DBAPPSecurity Co., Ltd.**
Fu Chunhui, Wang Xin

**Federal University of Rio Grande do Sul**
Edson Prestes

**Fuxi Institution**
Cheng Kai, Li Na

**University of Chinese Academy of Social Sciences**
Liu Xiaochun

**IQuilibriumAI**
Jimena Sofia Viveros Alvarez

**Lenovo Group**
Hu Yongqi

**Lingnan University of Hong Kong**
Yao Xin

**MiniMax**
Peng Tao, Shen Juncheng

**Moscow State Institute of International Relations**
Anna Abramova

**Nankai University**
Tao Feng

**Nokia Shanghai Bell Co., Ltd.**
Tao Tao, Wang Tong

**Peking University**
Sebastian Sunday Grève, Yang Yaodong

**People.cn Co., Ltd.**
Wang Fengqiong

**Renmin University of China**
Gong Xinqi, Liu Yongmou

**Shanghai AI Lab**
Qiao Yu

**Sina Weibo**
Wang Wei, Zhang Junlin

**Sri Lanka CERT**
Charuka Senal Damunupola, Nirosha Ananda

**Technical University of Munich**
Danil Kerimi

**Tencent**
Wang Mengyin, Wang Rong

**The Chinese University of Hong Kong**
Helen Meng, Zhang Jiji

**The University of Hong Kong**
George Chen

**Tsinghua University**
Tang Xinhua

**University of Erlangen-Nürnberg**
Vincent C. Müller

**University of Münster**
Bernd Holznagel

**University of Southampton**
Dame Wendy Hall

**World Internet Conference**
Liang Hao, Zhang Xueli

**Zhongguancun Laboratory**
Tan Zhixing

**ZTE Corporation**
Meng Wei

---

## Writing Group

Beijing Institute of AI Safety and Governance (Beijing-AISI)
Institute of Automation, Chinese Academy of Sciences
Beijing Key Laboratory of Safe AI and Superalignment
Zeng Yi, Wang Zhengqi, Lu Enmeng, Fan Jinyu, Huangfu Cunqing

World Internet Conference
Kang Yanrong, Han Kaiyu

University of Chinese Academy of Sciences
Cao Gongce, Chen Yu, Xie Jiawei, Han Zhengqiang, Guo Xiaoyang, Bao Aorigele, Wang Jin

## Contact us via:

research@wicinternet.org

# PREFACE

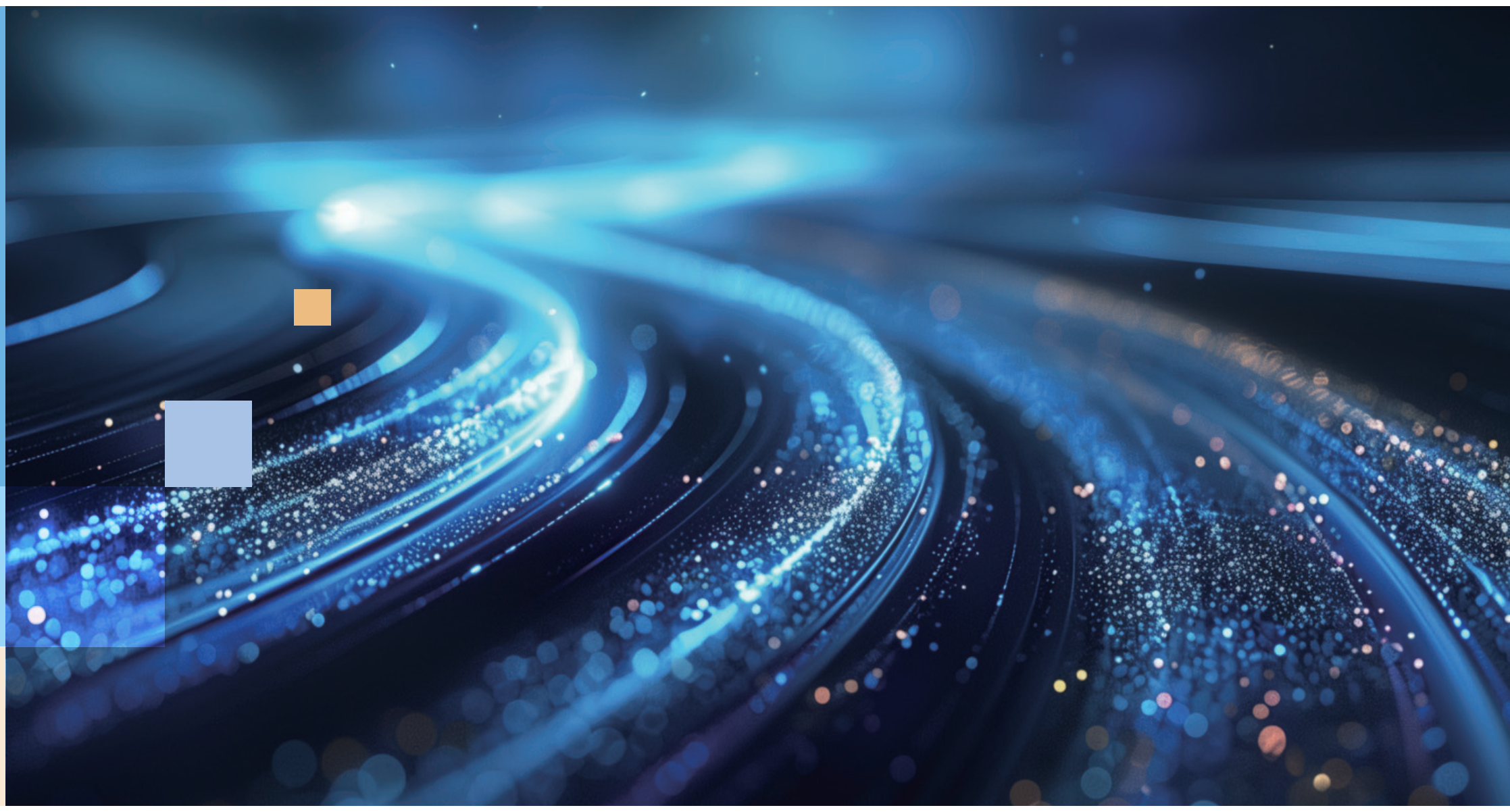

Artificial Intelligence (AI) is advancing at an exponential pace, profoundly reshaping the global landscape of science and technology, the economy, safety, and security. Alongside its widespread application, AI also brings about a wide range of unforeseen risks and complex challenges. Meanwhile, the global AI safety and governance system is showing signs of fragmentation. Differences in development levels, governance capacities, and governance priorities among countries have created obstacles to international cooperation. Given this context, exploring and constructing a robust global framework for AI safety and governance, with the aim of establishing widely accepted governance frameworks and standards at the earliest opportunity, has become a key task for ensuring the sustainable development of AI.

World Internet Conference (WIC), drawing upon its role as an international platform and leveraging the AI Safety and Governance Program under WIC Specialized Committee on AI, has joined forces with international organizations, leading think tanks, research institutes, professional associations, and industry experts around the world to advance this study. This effort aims to foster consensus, mutual trust, and collaboration among all parties through dialogue and cooperation, and to promote the use of AI in serving the common well-being and long-term safety of humanity.

This report provides a systematic review of the current exploration and practices in the global AI safety and governance system, identifies the key issues that need to be addressed, and draws from multilateral governance experiences from other global domains. Centered on mechanism design, it explores how to build a global AI safety and governance system that ensures safety and security, promotes inclusiveness, clarifies responsibilities and accountabilities, strengthens effective coordination, and ensures authoritative and efficient governance. Guided by the principles of multilateralism and the vision of building a community with a shared future for humankind, the report proposes institutional measures and policy recommendations to establish a United Nations-centered global AI safety and governance framework. It aims to serve as a reference for all stakeholders, foster broad consensus, and promote joint efforts to advance the construction and improvement of the global AI safety and governance system.

# Contents

— PART —

# 01

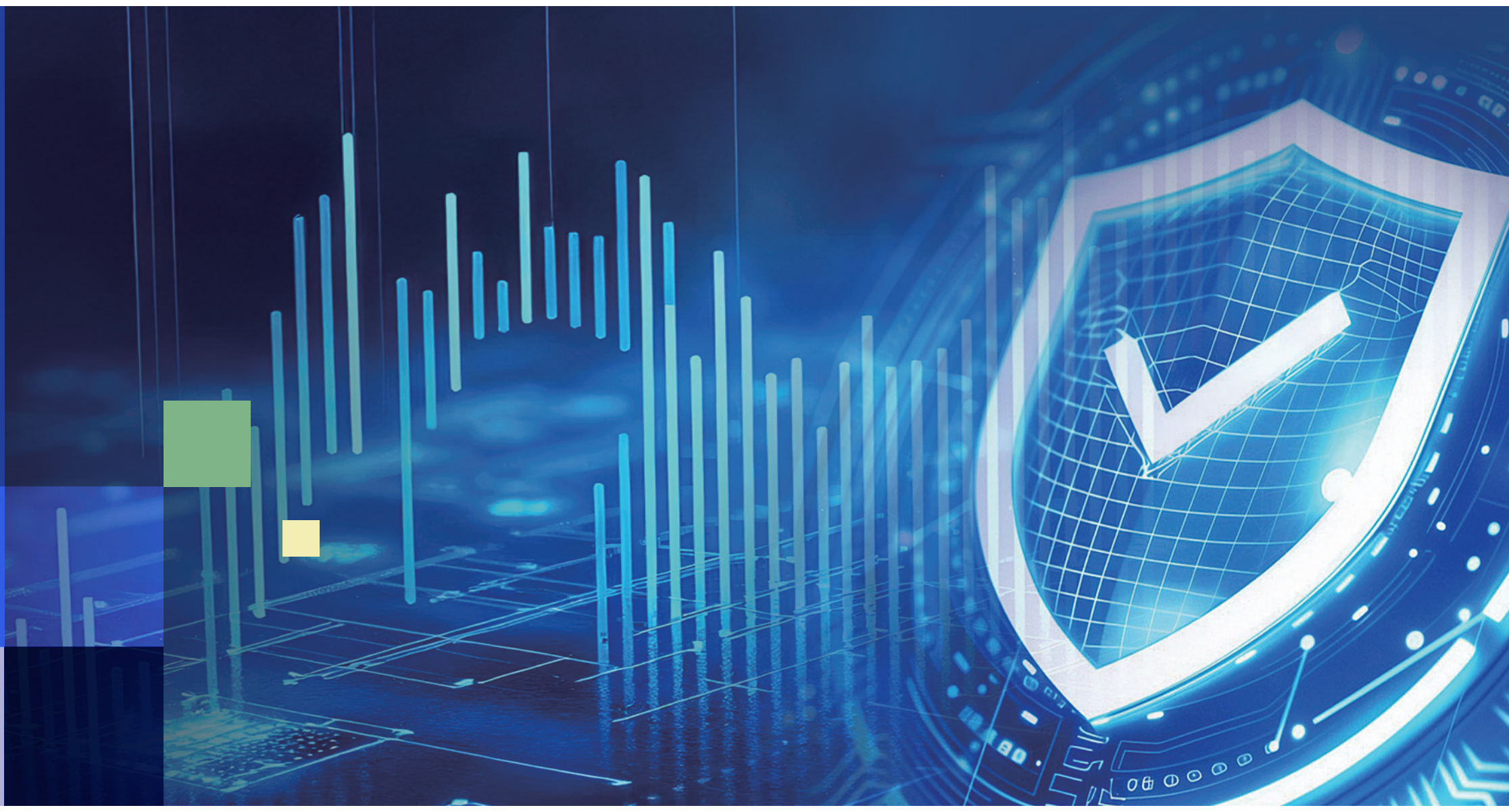

## Establishing a Global Framework for AI Safety and Governance

While artificial intelligence (AI) presents significant opportunities for development and transforms human production and lifestyles, it gives rise to cross-sectoral and multi-level systemic governance challenges, further intensifying the fragmentation and institutional fragility of the global governance landscape. **Promoting the formation of a global AI safety and governance framework that is both actionable and based on broad consensus has become a task of the era that humankind must face together.**

### (a) Addressing the Pressing Risks in Global AI Development and Application

**Currently, AI is advancing at an unprecedented pace, profoundly reshaping the global landscape of technology, economy, safety, and security.** Large-scale models have demonstrated reasoning capabilities in fields such as mathematics and programming that are on par with human experts; video generation and synthesis technologies have made a leap from conceptualization to highly realistic complex scene video creation within a short period; the development of embodied intelligence is also driving human-like interactions of general-purpose robots in complex physical environments. Meanwhile, AI has played a significant role in addressing major challenges in fields such as science, medicine, climate, energy, and transportation, creating substantial public value and social benefits.

**As AI technologies expand into broader and deeper application scenarios, governance challenges are becoming increasingly complex and exhibiting systemic spillover effects. In terms of inherent technological risks, the rapid iteration and capability leaps of algorithms have exposed security vulnerabilities such as insufficient interpretability, weak robustness, and adversarial fragility. At the application level, the widespread adoption of AI has brought issues such as computing power security, supply chain vulnerabilities, and cross-border technology**

**flows into sharp focus.** Regional and industrial critical systems are increasingly reliant on intelligent components, whose cascading effects may span across infrastructure sectors including energy, transportation, finance, and communications. Relevant statistics and risk assessments indicate a significant upward trend in malicious uses of AI and cyber attacks in recent years. The threat assessment published by the European Union Agency for Cybersecurity (ENISA) [1]shows that between July 2023 and June 2024, a total of 11,079 attack incidents were recorded, including 322 cross-border attacks affecting multiple EU member states. The primary targeted sectors were public administration (approximately 19%), transportation (11%), finance (9%), and digital infrastructure (8%). Simultaneously, deepfake activities powered by generative AI saw a 118% year-on-year increase in 2024[2]. The global economic losses resulting from related cybercrimes are projected to reach $10.5 trillion annually by 2025[3]. These risks have already expanded from the technological layer to social systems, characterized by cross-border and cross-sector transmission properties, posing severe challenges to existing governance frameworks. Against this backdrop, countries like China are exploring the establishment of systems such as content labeling and safety assessment for AI-generated synthetic content through laws and regulations, including the "Interim Measures for the Administration of Generative AI Services" [4]and the "Measures for Labeling of AI-Generated Synthetic Content"[5], in order to effectively mitigate related risks.

**Meanwhile, the continuous accumulation of AI risks, compounded by geopolitical factors, is further exacerbating global governance challenges.** In recent years, the number of AI-related risk incidents has continued to grown exponentially. As shown in Figure 1, between 2019 and 2024, the number of documented global AI risk incidents surged from approximately 400 to over 7,900—a nearly twentyfold increase. Among these, incidents involving robustness and digital security, human rights and privacy governance, transparency and accountability accounted for more than 60%, indicating that AI safety and ethical issues are emerging as global challenges. At the same time, some countries are pursuing strategic advantages through measures such as export controls and technological blockades[6]. Although strategic measures implemented under the guise of "technology control" have not effectively restricted the proliferation of AI, they may instead undermine the technological advantages of the policy-making countries[7]. This zero-sum game mindset is increasingly constraining space for international cooperation, weakening the capability and willingness of countries to participate in global AI safety and governance, and diminishing the effectiveness of the international community's collective response to systemic risks.

1. ENISA, "ENISA Threat Landscape 2024". Source: https://securitydelta.nl/media/com_hsd/report/690/document/ENISA-Threat-Landscape-2024.pdf
2. European Parliament, "Children and deepfakes". "2025 Identity Fraud Report". Source: https://www.europarl.europa.eu/RegData/etudes/BRIE/2025/775855/EPRS_BRI(2025)775855_EN.pdf
3. Entrust Cybersecurity Institute, "2025 Identity Fraud Report". Source: https://www.entrust.com/sites/default/files/documentation/reports/2025-identity-fraud-report.pdf
4. "Interim Measures for the Administration of Generative AI Services". Source: https://www.cac.gov.cn/2023-07/13/c_1690898327029107.htm
5. "Measures for Labeling of AI-Generated Synthetic Content". Source: https://www.gov.cn/zhengce/zhengceku/202503/content_7014286.htm
6. Understanding the Framework of AI Proliferation. Source: https://www.rand.org/pubs/perspectives/PEA3776-1.html
7. New AI Diffusion Export Control Rules Will Undermine U.S. AI Leadership. Source: https://www.brookings.edu/articles/the-new-ai-diffusion-export-control-rule-will-undermine-us-ai-leadership

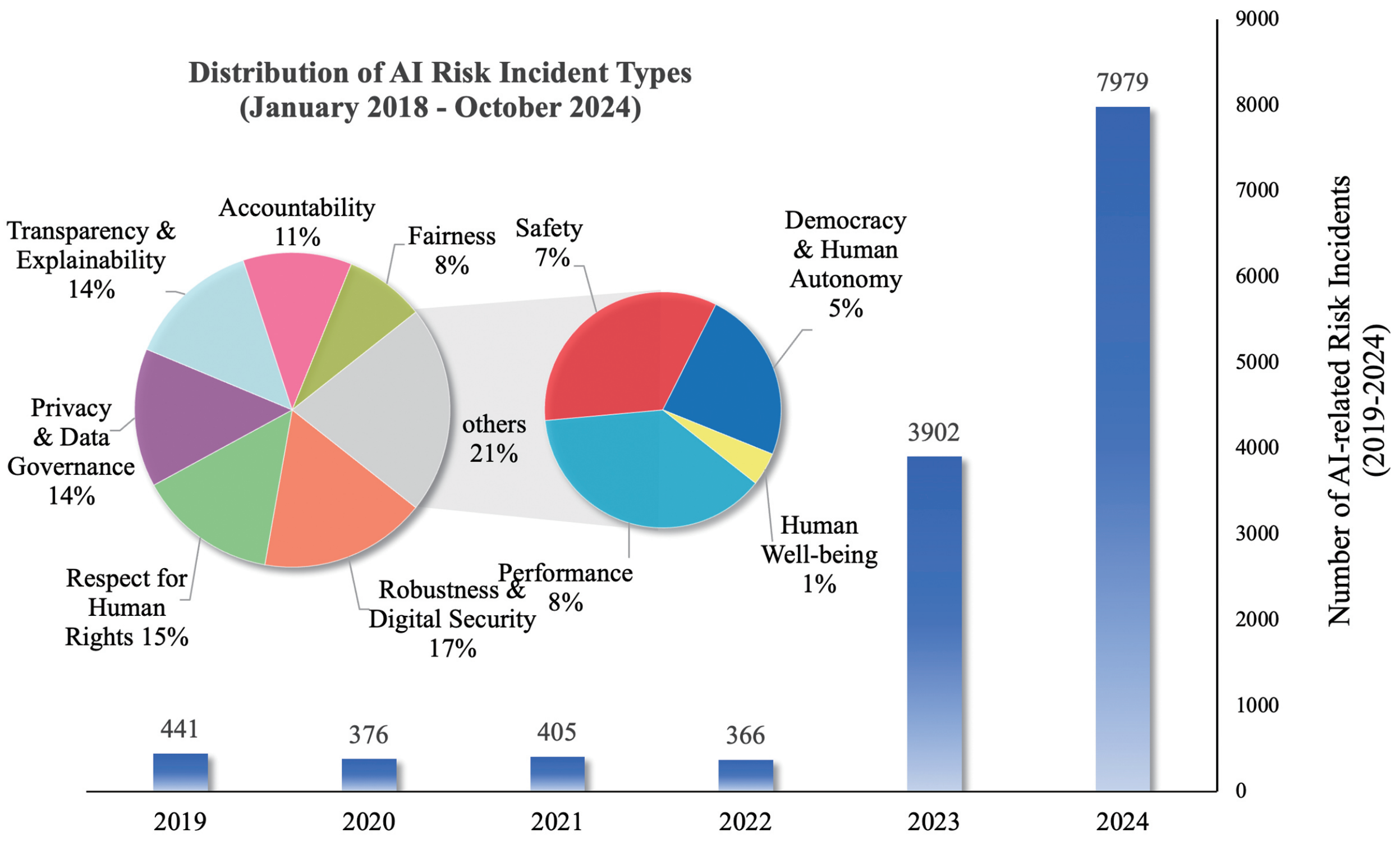

Figure 1: Trends in Global AI Risk Incident Volume and Distribution of Risk Types over Recent Years[8]

Faced with the reality of a significantly fragmented governance landscape and increasingly urgent safety and governance needs, **it has become imperative to establish a comprehensive global AI safety and governance framework from the perspective of safeguarding the common safety and long-term well-being of all humanity.** The cross-border nature of AI means its risks and impacts inherently transcend national boundaries—whether in data flows, algorithm deployment, or the application and proliferation of foundational models. Regulatory measures by individual countries or regions struggle to impose effective constraints, yet no nation can remain unaffected. Against the backdrop of continuously evolving technological risks with global spillover effects, **establishing a global AI safety and governance system—based on broad consensus, ensuring safety, security, and inclusiveness, with clear rights and responsibilities, coordinated effectively, and operating with authority and efficiency—has become a critical pathway to addressing systemic risks.** This is not only an inevitable requirement for technological development but also a crucial safeguard for maintaining collective safety in the global digital era.

## (b) Global Efforts and Practices in Building AI Safety and Governance Frameworks

**In recent years, global stakeholders have been jointly advancing the construction of a global AI safety and governance framework through multi-faceted and coordinated efforts.**

8. Source data from OECD AI Incidents Monitor (AIM), statistically analyzed by the "AI Governance InternationaL Evaluation Index 2025". Source: https://agile-index.ai/publications/2025w

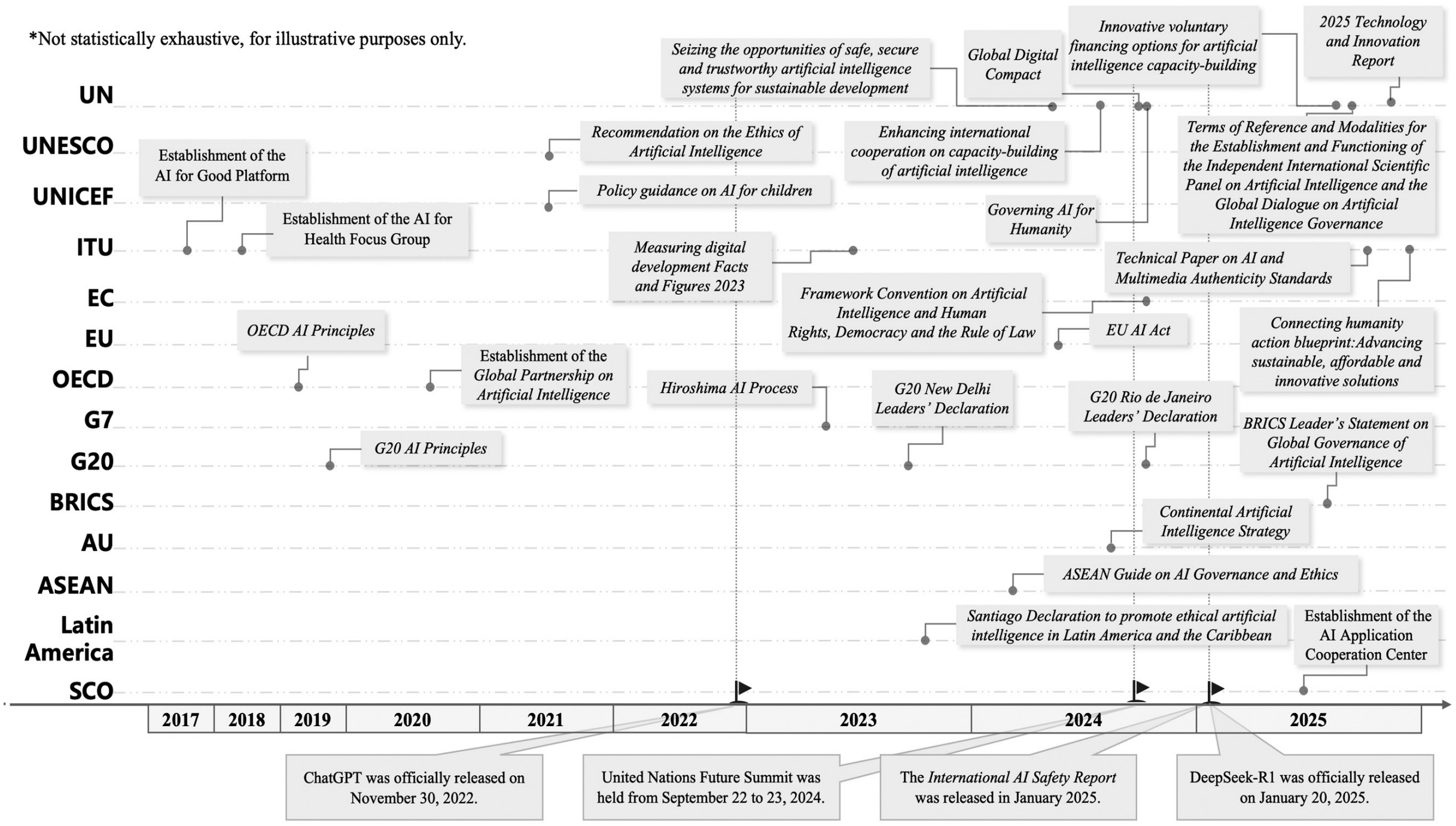

Figure 2: Recent Explorations by Various Inter-governmental Organizations in AI Safety and Governance

**The United Nations (UN) system** is actively constructing a global AI governance framework through measures such as establishing advisory bodies, convening meetings, issuing reports, and adopting resolutions by the General Assembly. **The UN Secretary-General** established the High-Level Advisory Body on Artificial Intelligence [9], and released both the interim[10] and final [11]reports titled "Governing AI for Humanity", which analyze international AI governance issues and put forward recommendations. **The United Nations General Assembly (UNGA)** has successively adopted resolutions titled "Seizing the opportunities of safe, secure and trustworthy artificial intelligence systems for sustainable development" [12]and "Enhancing international cooperation on capacity-building of artificial intelligence"[13]. It convened the Future Summit, adopted the "Pact for the Future"[14] which incorporated the "Global Digital Compact"[15], and adopted subsequent resolutions including the "Terms of Reference and Modalities for the Establishment and Functioning of the Independent International Scientific Panel on Artificial Intelligence and the Global Dialogue on Artificial Intelligence Governance" [16], thereby continuously advancing the implementation of the global AI governance agenda. Furthermore,

9. Secretary-General forms high-level advisory body with 39 experts from around the world to discuss AI governance. Source: https://www.un.org/en/ai-advisory-body/about
10. United Nations, "Governing AI for Humanity" Interim Report. Source: https://www.un.org/sites/un2.un.org/files/ai_advisory_body_interim_report.pdf
11. United Nations, "Governing AI for Humanity" Final Report. Source: https://www.un.org/sites/un2.un.org/files/governing_ai_for_humanity_final_report_en.pdf
12. United Nations, "Seizing the opportunities of safe, secure and trustworthy artificial intelligence systems for sustainable development". Source: https://docs.un.org/en/A/78/L.49
13. United Nations, "Enhancing international cooperation on capacity-building of artificial intelligence". Source:https://docs.un.org/en/A/RES/78/311
14. UN Future Summit, "The Pact for the Future". Source: https://docs.un.org/en/A/RES/79/1
15. UN Future Summit, "Global Digital Compact". Source: https://www.un.org/en/summit-of-the-future/global-digital-compact
16. The United Nations General Assembly decides to establish an independent international panel on AI. Source: https://docs.un.org/zh/A/RES/79/325

the UN Secretary-General submitted to the General Assembly the report titled "Innovative voluntary financing options for artificial intelligence capacity-building"[17], further exploring relevant financing mechanisms and funding solutions to implement AI capacity-building. **United Nations Educational, Scientific and Cultural Organization (UNESCO)** issued the "Recommendation on the Ethics of Artificial Intelligence"[18], along with complementary tools including the "Readiness Assessment Methodology" [19]and the "Ethical Impact Assessment"[20], providing normative guidance for Member States and stakeholders. **International Telecommunication Union (ITU)** convened the AI for Good Global Summit[21], focusing on innovative applications of artificial intelligence and aiming to promote solutions to global challenges. **United Nations Children's Fund (UNICEF)** launched "Policy guidance on AI for children" [22], which outlines the principles of protection, empowerment, and friendliness that should be followed in developing child-friendly AI.

**At the national level, countries are actively constructing governance frameworks by issuing strategic documents and action initiatives.** For example, the **United States,** through its National Institute of Standards and Technology (NIST), published the "Artificial Intelligence Risk Management Framework"[23], providing public and private sectors with methodological tools to identify and manage potential risks throughout the AI lifecycle; has also promoted the establishment of the International Network of AI Safety Institutes[24], and is conducting international cooperation on frontier model joint testing and safety evaluation. **Singapore** has issued the "Model AI Governance Framework (Second Edition)" [25] and the "Model AI Governance Framework for Generative AI: Fostering a Trusted Ecosystem"[26], offering policy guidance and practical references for government, industry, and research institutions in the development and application of AI. **Saudi Arabia** has issued the "AI Ethics Principles"[27], emphasizing the safety and reliability of AI systems. **China** has successively released the "Global AI Governance Initiative"[28], "AI Safety Governance Framework (1.0, 2.0)"[29], "Global AI Governance Action Plan"[30], advocating the principle of balanced development and safety, proposing fun-

17. United Nations, "Innovative voluntary financing options for artificial intelligence capacity-building". Source: https://digitallibrary.un.org/record/4085951?ln=en&v=pdf#files
18. UNESCO, "Recommendation on the Ethics of Artificial Intelligence". Source: https://unesdoc.unesco.org/ark:/48223/pf0000380455
19. UNESCO, "Readiness Assessment Methodology (RAM)". Source: https://www.unesco.org/ethics-ai/en/ram?hub=32618
20. UNESCO, "Ethical Impact Assessment (EIA)". Source: https://www.unesco.org/ethics-ai/en/eia?hub=32618
21. AI for Good Global Summit organized by the International Telecommunication Union (ITU), is convened in collaboration with the Government of Switzerland and brings together over 50 United Nations partners. Source:https://aiforgood.itu.int/#
22. UNICEF, "Policy guidance on AI for children". Source: https://www.unicef.org/innocenti/reports/policy-guidance-ai-children
23. The U.S. National Institute of Standards and Technology, "Artificial Intelligence Risk Management Framework". Source: https://www.nist.gov/itl/ai-risk-management-framework
24. U.S. Department of Commerce & U.S. Department of State Launch the International Network of AI Safety Institutes at Inaugural Convening in San Francisco. Source: https://www.nist.gov/news-events/news/2024/11/fact-sheet-us-department-commerce-us-department-state-launch-international
25. Singapore, "Model AI Governance Framework (Second Edition)". Source: https://www.pdpc.gov.sg/-/media/Files/PDPC/PDF-Files/Resource-for-Organisation/AI/SGModelAIGovFramework2.pdf
26. Singapore, "Model AI Governance Framework for Generative AI: Fostering a Trusted Ecosystem". Source: https://aiverifyfoundation.sg/wp-content/uploads/2024/05/Model-AI-Governance-Framework-for-Generative-AI-May-2024-1-1.pdf
27. Saudi Arabia's AI Ethics Principles. Source: https://sdaia.gov.sa/en/SDAIA/about/Documents/ai-principles.pdf
28. China, "Global AI Governance Initiative". Source: https://www.mfa.gov.cn/eng/zy/gb/202405/t20240531_11367503.html
29. China has released and updated the "AI Safety Governance Framework". Source: https://www.cac.gov.cn/2024-09/09/c_1727567886199789.htm, https://www.cac.gov.cn/2025-09/15/c_1759653448369123.htm
30. China, "Global AI Governance Action Plan". Source: https://www.fmprc.gov.cn/mfa_eng/xw/zyxw/202507/t20250729_11679232.html

damental guidelines for trustworthy AI to address the risks of technological loss of control, and introducing systematic implementation measures focusing on technical supervision, industrial promotion, and international cooperation. China has also proposed the establishment of the World Artificial Intelligence Cooperation Organization[31], contributing to global AI governance from a Chinese perspective.

**At the regional level, Europe** has taken the lead in establishing a relatively comprehensive regulatory and legal framework. The European Commission issued the "Coordinated Plan on Artificial Intelligence" [32]and the "White Paper on Artificial Intelligence - A European approach to excellence and trust"[33]. The European Parliament and the Council of the European Union adopted the "Artificial Intelligence Act"[34], establishing the world's first comprehensive legal framework to regulate AI. The Council of Europe introduced the first legally binding international treaty on AI: "Council of Europe Framework Convention on Artificial Intelligence and Human Rights, Democracy and the Rule of Law" [35], which aims to ensure that activities throughout the AI system lifecycle are fully consistent with human rights, democracy, and the rule of law, while fostering technological innovation. **In the Arab region**, the League of Arab States (LAS), through its Arab AI Working Group and in close collaboration with regional specialized organizations such as the Arab ICT Organization (AICTO), has successively released and adopted the "Arab AI Strategy"[36] and the "Arab AI Ethics Charter"[37], advancing regional coherence in the development of AI governance and ethical frameworks. **In Africa,** the African Union released the "Continental Artificial Intelligence Strategy"[38], establishing a regional-level policy framework and directions for action. Building on this, multiple African countries signed "The Africa Declaration on Artificial Intelligence"[39], fostering a continent-wide political consensus and a shared vision for coordinated action. **In Southeast Asia,** the Association of Southeast Asian Nations (ASEAN) released the "ASEAN Guide on AI Governance and Ethics" [40]and subsequently published the "ASEAN Guide on AI Governance and Ethics–Generative AI"[41], which outlines the risks associated with generative AI and recommends policy measures to ensure the responsible development and use of AI across the region. **In Latin America,** representatives from Latin American and Caribbean countries reached a consensus and issued the "Santiago Declaration: To Promote Ethical Artificial Intelligence in Latin America and the Ca-

31. China initiated the establishment of the World Artificial Intelligence Cooperation Organization (WAICO). Source: https://www.gov.cn/yaowen/liebiao/202507/content_7033957.htm
32. European Commission, "Coordinated Plan on Artificial Intelligence". Source: https://digital-strategy.ec.europa.eu/en/policies/plan-ai
33. European Commission, "White Paper on Artificial Intelligence". Source: https://digital-strategy.ec.europa.eu/en/library/white-paper-artificial-intelligence-public-consultation-towards-european-approach-excellence-and
34. European Commission, "Artificial Intelligence Act". Source: https://artificialintelligenceact.eu/the-act/
35. Council of Europe, "Framework Convention on Artificial Intelligence and Human Rights, Democracy and the Rule of Law". Source: https://www.coe.int/en/web/portal/-/council-of-europe-opens-first-ever-global-treaty-on-ai-for-signature
36. "Arab AI Strategy". Source: https://www.aicto.org/publications/studies/#flipbook-df_9695/1/
37. "Arab AI Ethics Charter". Source: https://www.aicto.org/publications/studies/#flipbook-df_9686/1/
38. African Union, "Continental Artificial Intelligence Strategy". Source: https://au.int/en/documents/20240809/continental-artificial-intelligence-strategy
39. Global AI Summit on Africa, "The Africa Declaration on Artificial Intelligence". Source: https://c4ir.rw/docs/Africa%20Declaration%20on%20Artificial%20Intelligence-FINAL-31-March-2025.pdf
40. Association of Southeast Asian Nations, "ASEAN Guide on AI Governance and Ethics". Source: https://asean.org/book/asean-guide-on-ai-governance-and-ethics/
41. Association of Southeast Asian Nations, "ASEAN Guide on AI Governance and Ethics – Generative AI". Source: https://mekongdataprotection.org/asean-guide-on-ai-governance-and-ethics-generative-ai

ribbean" [42]and the "Declaration of Montevideo: For the Construction of a Regional Approach on the Governance of Artificial Intelligence and Its Impacts on Our Society"[43], encouraging regional coordination and high-level dialogue. These measures reflect the active efforts of different regions in the development and governance of AI, emphasizing regional coordination and ethical responsibility, and calling for the promotion of a fair and inclusive governance framework on a global scale.

At the international organizations level, the **Group of Twenty (G20)** 2023 New Delhi Summit released the "G20 New Delhi Leaders' Declaration"[44], which further affirmed the commitment to ensuring that AI serves the good of all humanity. The 2024 Rio de Janeiro Summit issued the "G20 Rio de Janeiro Leaders' Declaration" [45], calling for the promotion of AI governance conducive to innovation, enhanced cooperation, and the empowerment of sustainable development. The **Group of Seven (G7)** launched the "G7 Leaders' Statement on the Hiroshima AI Process"[46], proposing the establishment of a common governance framework and emphasizing cooperation on safety, transparency, and responsible application. The **Organisation for Economic Co-operation and Development (OECD)** issued the "OECD AI Principles" [47]and launched the Global Partnership on Artificial Intelligence (GPAI)[48] to advance international collaboration in AI policy practice, technological research, and capacity building. The Asia-Pacific Economic Cooperation (APEC) released the "APEC Artificial Intelligence Initiative (2026-2030)"[49], committing to promoting secure, accessible and reliable AI ecosystems to achieve resilient and inclusive economic growth. The **BRICS countries** released the "Rio de Janeiro Declaration" of the 17th BRICS Summit[50] and the "BRICS Leaders' Statement on The Global Governance of Artificial Intelligence"[51], actively advocating for an international AI governance framework and cooperation platform oriented toward inclusiveness, fairness, and sustainable development. The Shanghai Cooperation Organization (SCO) issued the "Statement on Further Deepening International Cooperation in Artificial Intelligence" during the 2025 Tianjin Summit[52], advocating for enhanced cooperation in areas such as AI infrastructure, talent cultivation, and investment. It also called for the development of dialogue partnership mechanisms in the field

42. "Santiago Declaration: To Promote Ethical Artificial Intelligence in Latin America and the Caribbean". Source: https://minciencia.gob.cl/uploads/filer_public/40/2a/402a35a0-1222-4dab-b090-5c81bbf34237/declaracion_de_santiago.pdf
43. "Declaration of Montevideo: For the Construction of a Regional Approach on the Governance of Artificial Intelligence and Its Impacts on Our Society". Source: https://www.gub.uy/agencia-gobierno-electronico-sociedad-informacion-conocimiento/sites/agencia-gobierno-electronico-sociedad-informacion-conocimiento/files/documentos/noticias/EN%20-%20Montevideo%20Declaration%20approved.pdf
44. G20, "G20 New Delhi Leaders' Declaration". Source: https://www.consilium.europa.eu/media/66739/g20-new-delhi-leaders-declaration.pdf
45. G20, "G20 Rio de Janeiro Leaders' Declaration". Source: https://g20.org/wp-content/uploads/2024/11/G20-Rio-de-Janeiro-Leaders-Declaration-EN.pdf
46. G7, "G7 Leaders' Statement on the Hiroshima AI Process". Source: https://g7g20-documents.org/database/document/2023-g7-japan-leaders-leaders-language-g7-leaders-statement-on-the-hiroshima-ai-process
47. OECD, "OECD AI principles". Source: https://www.oecd.org/en/topics/sub-issues/ai-principles.html
48. About the Global Partnership on Artificial Intelligence (GPAI). Source: https://oecd.ai/en/about/about-gpai
49. APEC, "APEC Artificial Intelligence Initiative (2026-2030)". Source: https://www.apec.org/meeting-papers/leaders-declarations/2025/2025-apec-leaders--gyeongju-declaration/apec-artificial-intelligence-(ai)-initiative-(2026-2030)
50. BRICS, the "Rio de Janeiro Declaration" of the 17th BRICS Summit. Source: http://brics.br/en/documents/presidency-documents/250705-brics-leaders-declaration-en.pdf
51. BRICS, "BRICS Leaders' Statement On The Global Governance Of Artificial Intelligence". Source: http://brics.br/en/documents/presidency-documents/250706_brics_ggai_declarationfinal.pdf
52. Statement by the Council of Heads of State of the Shanghai Cooperation Organization Member States on Further Deepening International Cooperation in Artificial Intelligence. Source: https://chn.sectsco.org/20250901/1969229.html

of AI, focusing on the sustainable development of the AI industry and addressing potential risks and challenges posed by AI. **World Internet Conference** released documents and reports such as the "Developing Responsible Generative Artificial Intelligence Research Report and Consensus"[53] and "Governing AI for Good and for All - Empowering Global Sustainable Development"[54]. These documents propose actively promoting and steadily advancing the development of generative AI while bridging the digital and AI divide through inclusive and equitable AI development and governance.

**Major international summits** have also fostered global consensus on key issues by establishing principles and building mechanisms. For example, since 2023, the series of AI Summits including the **AI Safety Summit,** the **AI Seoul Summit,** and the **AI Action Summit** has produced a number of voluntary outcomes, such as the "Bletchley Declaration"[55], the "Seoul Ministerial Statement for advancing AI safety, innovation and inclusivity: AI Seoul Summit 2024"[56], the "Frontier AI Safety Commitments"[57], and the "Statement On Inclusive And Sustainable Artificial Intelligence For People And The Planet"[58]. The series of **Summits on Responsible AI in the Military Domain (REAIM)** has, since 2023, signed and issued the "REAIM Call to Action"[59] and the "REAIM Blueprint for Action"[60], and released the report "Responsible by Design: Strategic Guidance Report on the Risks, Opportunities, and Governance of AI in the Military Domain"[61], reflecting the shared concern of the international community over the responsible use of AI in the military domain.

Furthermore, both **academia** and **industry** are actively exploring risk response mechanisms and promoting self-regulation and autonomy. Through theoretical construction and shaping international consensus, **academia** enhances the understanding and awareness of risks associated with frontier AI. For example, the International Dialogue on AI Safety (IDAIS)[62] fosters scientific consensus on addressing extreme risks of AI by organizing exchanges among top global experts and advocates the establishment of international safety standards and governance frameworks. The Center for AI Safety (CAIS) collaborates with scholars and policymakers to jointly sign the "Statement on AI Risk"[63], promoting the formation of a global

53. World Internet Conference, "Developing Responsible Generative Artificial Intelligence (AI) Research Paper and Consensus". Source: https://www.wicinternet.org/pdf/DevelopingResponsibleGenerativeArtificialIntelligenceResearchReportandConsensus.pdf
54. World Internet Conference, "Governing AI for Good and for All-Empowering Global Sustainable Development". Source: https://wicinternet.org/2025-04/13/c_1081925.htm
55. AI Safety Summit 2023, "The Bletchley Declaration". Source: https://www.gov.uk/government/publications/ai-safety-summit-2023-the-bletchley-declaration/the-bletchley-declaration-by-countries-attending-the-ai-safety-summit-1-2-november-2023
56. AI Seoul Summit 2024, "Seoul Ministerial Statement for advancing AI safety, innovation and inclusivity: AI Seoul Summit 2024". Source: https://www.gov.uk/government/publications/seoul-ministerial-statement-for-advancing-ai-safety-innovation-and-inclusivity-ai-seoul-summit-2024
57. AI Seoul Summit 2024, "Frontier AI Safety Commitments". Source:https://www.gov.uk/government/publications/frontier-ai-safety-commitments-ai-seoul-summit-2024
58. AI Action Summit "Statement On Inclusive And Sustainable Artificial Intelligence For People And The Planet". Source: https://www.elysee.fr/en/emmanuel-macron/2025/02/11/statement-on-inclusive-and-sustainable-artificial-intelligence-for-people-and-the-planet
59. REAIM 2023, "REAIM Call to Action". Source: https://www.government.nl/documents/publications/2023/02/16/reaim-2023-call-to-action
60. REAIM 2024, "REAIM Blueprint for Action". Source: https://thereadable.co/reaim-blueprint-for-responsible-ai-use-military/
61. "Responsible by Design: Strategic Guidance Report on the Risks, Opportunities, and Governance of AI in the Military Domain". Source: https://hcss.nl/wp-content/uploads/2025/09/GC-REAIM-Strategic-Guidance-Report-Final-WEB.pdf
62. International Dialogues on AI Safety (IDAIS) brings together senior scientists worldwide to jointly address extreme risks posed by AI. Source: https://idais.ai/dialogues
63. Center for AI Safety (CAIS), "Statement on AI Risk". Source: https://aistatement.com/

governance collaboration. **The industry** is promoting the prudent deployment of high-risk AI systems and fostering industry consensus by developing autonomous safety frameworks. Many leading AI companies have successively released their respective risk control frameworks, aiming to ensure the responsible deployment of AI systems and to link the capabilities of frontier models to their potential risks. These include but are not limited to **Anthropic**'s "Responsible Scaling Policy"[64], **OpenAI**'s "Preparedness Framework" [65], **Google DeepMind**'s "Frontier Safety Framework"[66], and **Meta**'s "Outcome-Focused Frontier AI Framework"[67]. The **China AI Industry Alliance (AIIA)** has issued the "AI Safety Commitment" [68], which focuses on compliance, transparency, and controllability in AI research, development, and application, advocating for enterprises to proactively fulfill their primary safety responsibilities.

## (c) Key Challenges for a Global Framework on AI Safety and Governance

Although global stakeholders have made extensive efforts and achieved notable progress in areas such as AI safety, ethical standards, capacity building, and cross-border cooperation, **current global AI governance still faces structural limitations in addressing the unprecedented risks and challenges arising from AI development, as reflected in the following aspects. First,** the highly unpredictable nature of AI technological breakthroughs and application pathways, coupled with the complexity and diversity of risks, makes it difficult to implement forward-looking governance measures and maintain dynamic adaptability. An agile and coordinated global response mechanism has yet to be established. **Second,** significant disparities in AI development levels and governance models among countries have led to markedly different capacities for participation and institutional demands, posing challenges to building international consensus. **Third,** the diversity of AI stakeholders, without clear accountability mechanisms and effective supervision and implementation, hinders coordinated and robust global action. **Fourth,** current disruptive factors such as decoupling and supply chain fragmentation, driven by geopolitical tensions, ideological differences, and economic interests, are undermining the global capacity to collectively address AI risks.

**Therefore, guided by the vision of a shared future for humanity, establishing a global AI safety and governance framework with broad international consensus must focus on addressing the following key issues:**

**Firstly,** how to effectively mitigate the uncertainties of AI technical safety risks and their derivatives, ensuring global preparedness and adequate response capabilities against potential major threats to maintain security and controllability.

**Secondly,** how to balance the considerable disparities among nations in development stages,

64. Anthropic, "Responsible Scaling Policy". Source: https://www.anthropic.com/responsible-scaling-policy
65. OpenAI, "Preparedness Framework". Source: https://cdn.openai.com/pdf/18a02b5d-6b67-4cec-ab64-68cdfbddebcd/preparedness-framework-v2.pdf
66. GoogleDeepMind, "Frontier Safety Framework". Source: https://storage.googleapis.com/deepmind-media/DeepMind.com/Blog/introducing-the-frontier-safety-framework/fsf-technical-report.pdf
67. Meta, "Frontier AI Framework". Source: https://ai.meta.com/static-resource/meta-frontier-ai-framework/
68. China AI Industry Alliance (AIIA) initiated the "AI Safety Commitments". Source: https://aihub.caict.ac.cn/ai_security_and_safety_commitments

institutions, cultures, and governance capacities and objectives, in order to maximize common development and ensure AI benefits are shared inclusively.

**Thirdly,** how to coordinate complex interactions among diverse AI stakeholders. The goal is to harness their initiative and strengths to build synergy, thereby fostering a collaborative and efficient multi-stakeholder governance framework and preventing disorder that creates governance gaps.

**Fourthly,** how to overcome divisions caused by geopolitics and ideology to progressively build an authoritative and effective global AI safety and governance regime that can operate sustainably for the long term.

**In summary, the four key issues outlined above are interconnected and mutually reinforcing, collectively forming the foundational pillars of the global AI governance framework.** Ensuring safety stands as the foremost objective in building this governance system; ensuring inclusiveness serves as the cornerstone for establishing an equitable governance structure and constitutes the prerequisite for joint consultation, co-construction, and shared benefits among all countries; clarifying responsibilities and accountabilities provides the essential support for fostering proactive engagement by all stakeholders and ensuring that the governance system operates in a coordinated and effective manner; and adopting a future-oriented perspective offers the essential value guidance to safeguard the governance system from obstructive factors and short-sighted actions, thereby uniting global efforts toward a robust, resilient and sustainable AI safety and governance ecosystem.

## (d) Insights from Global Multilateral Governance Frameworks in Other Sectors

**In critical domains concerning humanity's collective safety, cross-border collaboration, and risk response—including nuclear safety, climate change, international trade, and public health—the international community has established global governance frameworks through multilateral agreements, specialized agencies, and dispute settlement mechanisms. These practices provide important institutional references and empirical insights for further constructing and refining the global security and governance framework for AI.**

**In international nuclear safety,** the International Atomic Energy Agency (IAEA) has established a mechanism combining rigorous inspections[69]with technical assistance[70] to address potential risks of nuclear technology. By establishing authoritative regulatory systems[71], supporting compliance monitoring and verification,

69. Treaty on the Non-Proliferation of Nuclear Weapons (NPT), preventing the proliferation of nuclear weapons, promoting nuclear disarmament, and facilitating the peaceful use of nuclear energy. Source: https://disarmament.unoda.org/en/our-work/weapons-mass-destruction/nuclear-weapons/treaty-non-proliferation-nuclear-weapons
70. "Convention on Assistance in the Case of a Nuclear Accident or Radiological Emergency" provides an international framework for relevant countries to request and provide assistance. Source: https://www.iaea.org/topics/nuclear-safety-conventions/convention-assistance-case-nuclear-accident-or-radiological-emergency
71. Convention on Early Notification of a Nuclear Accident, which requires State Parties to provide immediate notification in the event of a nuclear accident that may have radiological safety implications for other countries. Source: https://www.iaea.org/topics/nuclear-safety-conventions/convention-early-notification-nuclear-accident

and capacity building[72], the IAEA has fostered global collaboration and trust in nuclear safety, developing an integrated governance model balancing enforceability and collaboration. This can provide valuable insights for AI governance to establish credible compliance mechanisms and enhance risk prevention and emergency response capabilities.

**In climate change governance,** the Intergovernmental Panel on Climate Change (IPCC)[73] has created a mechanism integrating interdisciplinary scientific assessments[74] with policy-relevant recommendations[75] to address uncertainties in global climate change. Through iterative cross-disciplinary scientific evaluations, expert collaboration, and data-sharing platforms[76], the IPCC has facilitated global climate consensus and coordinated response actions[77], shaping an adaptive governance model where scientific evidence informs policy and frameworks evolve dynamically. This provides a blueprint for AI governance to develop science-based risk assessment systems amid high uncertainty and complex, cross-domain risks.

**In international trade,** the World Trade Organization (WTO)[78] has established a rule-based dispute settlement mechanism with binding arbitration to resolve trade frictions and rule conflicts among members[79]. Through uniform rule enforcement, arbitral awards, and balancing multi-stakeholder interests, the WTO has enabled efficient, fair, and orderly cross-border trade[80], establishing a rules-based governance model that balances diverse stakeholder interests through consensus-building[81]. This offers insights for AI governance to design transparent, impartial, and enforceable dispute resolution mechanisms.

**In public health,** the World Health Organization (WHO) has instituted a legally binding institutional framework under the International Health Regulations (IHR) covering epidemic reporting, cross-border source tracing, emergency

72. "Convention on Nuclear Safety", which ensures the safe operation of civilian nuclear power plants and promotes continuous improvement of nuclear safety through peer reviews. Source: https://www.iaea.org/topics/nuclear-safety-conventions/convention-nuclear-safety
73. "United Nations Framework Convention on Climate Change (UNFCCC)", the foundational legal framework for global climate change response. Source: https://www.un.org/climatesecuritymechanism/en/united-nations-framework-convention-climate-change-unfccc-and-climate-peace-and-security
74. A Comprehensive and Balanced Assessment of the State of Knowledge on IPCC Climate Change-Related Topics. Source: https://www.ipcc.ch/about/preparingreports/
75. IPCC, Summary for Policymakers. Source: https://www.ipcc.ch/report/ar6/syr/summary-for-policymakers/
76. IPCC Data. Source: https://www.ipcc.ch/data/
77. "Kyoto Protocol", which sets legally binding quantitative emission reduction and limitation targets. Source: https://unfccc.int/kyoto_protocol. Paris Agreement, which promotes the establishment of a global framework for climate action. Source: https://unfccc.int/process-and-meetings/the-paris-agreement
78. "Marrakesh Agreement Establishing the World Trade Organization", which established the World Trade Organization and provided the legal basis for its functions of administering multilateral trade rules, providing a forum for negotiations, and settling disputes. Source: https://www.wto.org/english/res_e/booksp_e/agrmntseries1_wto_e.pdf
79. WTO, Annex 2 of the WTO Agreement: "Understanding on rules and procedures governing the settlement of disputes". Source: https://www.wto.org/english/tratop_e/dispu_e/dsu_e.htm
80. Understanding The WTO: Settling Disputes. Source: https://www.wto.org/english/thewto_e/whatis_e/tif_e/disp1_e.htm
81. "General Agreement on Tariffs and Trade (GATT)", which established fundamental rules for trade in goods, including non-discrimination, reciprocity, and tariff concessions. Source: https://www.wto.org/english/res_e/publications_e/ai17_e/gatt1994_e.htm
"General Agreement on Trade in Services (GATS)", aimed at extending multilateral trade rules to the services sector and establishing a transparent, predictable, and progressively liberalized legal framework for it. Source: https://www.wto.org/english/tratop_e/serv_e/gatsintr_e.htm
"Agreement on Trade-Related Aspects of Intellectual Property Rights (TRIPS)", which set minimum international standards for intellectual property regulation and integrated intellectual property into the multilateral trading system to address trade-related intellectual property issues. Source: https://www.wto.org/english/tratop_e/trips_e/ta_modules_e.htm

quarantine, and resource allocation to address transnational pandemic risks[82]. Through standardized information notification, resource coordination, and emergency response protocols[83], WHO has enhanced the collaborative prevention and crisis response capabilities of the global public health system[84], developing a cross-border linkage and rapid response governance model. This provides a valuable reference for AI governance to establish efficient information-sharing and risk early-warning mechanisms, as well as to build rapid multilateral coordination and intervention capabilities in the face of sudden risk diffusion.

**In global aviation and maritime transport,** the International Civil Aviation Organization (ICAO) [85]and International Maritime Organization (IMO) [86]have developed holistic governance systems[87] integrating standardization, risk-based oversight, incident investigation, and compliance auditing to address transnational safety and interoperability challenges. By harmonizing technical standards, strengthening cross-system risk management, and conducting continuous safety assessments, they have enhanced the safety and interoperability of international transport systems, evolving into a governance model featuring standard harmonization and adaptive regulation. This can help guide AI governance in establishing unified standards and liability systems for infrastructure safety, cross-platform interoperability, and accident traceability, ensuring safe operation and transparent accountability in high-risk scenarios.

While the global governance systems in the aforementioned fields have accumulated rich experience in institutional design, compliance monitoring, and risk management, AI technology possesses unprecedented complex characteristics—including generative capacity, autonomy, evolvability, and cross-domain penetration—that result in risks with broader spillover effects, longer impact chains, and more difficult-to-define governance boundaries. **Consequently, global AI safety and governance cannot simply replicate existing institutional models. Instead, it must build upon the foundations of established mechanisms to develop new institutional approaches capable of ad-**

82. WHO, “The International Health Regulations (IHR)”—a legally binding global framework for public health security designed to help countries prevent and respond to public health emergencies of international concern. Source: https://www.who.int/health-topics/international-health-regulations#tab=tab_1
83. IHR Emergency Committees. Source: https://www.who.int/teams/ihr/ihr-emergency-committees
84. “WHO Framework Convention on Tobacco Control”. The first global public health treaty negotiated under the auspices of the World Health Organization, which aims to address the global tobacco epidemic by reducing tobacco consumption and supply. Source: https://wkc.who.int/resources/publications/i/item/9241591013
85. “Convention on International Civil Aviation (Chicago Convention)”, which establishes the fundamental legal framework and principles for international civil aviation activities and created the International Civil Aviation Organization (ICAO). Source: https://www.icao.int/convention-international-civil-aviation-doc-7300
86. “United Nations Convention on the Law of the Sea (UNCLOS)”, an international treaty often referred to as the “Constitution for the Oceans,” which systematically establishes a legal framework and order for all oceans and maritime activities. Source: https://www.imo.org/en/ourwork/legal/pages/unitednationsconventiononthelawofthesea.aspx
87. “International Convention for the Safety of Life at Sea (SOLAS)”, an international treaty under the auspices of the International Maritime Organization (IMO) that aims to establish uniform standards for the construction, equipment, and operation of merchant ships to ensure safety. Source: https://www.imo.org/en/knowledgecentre/conferencesmeetings/pages/solas.aspx
“York-Antwerp Rules”, a set of maritime customary rules with a history of several centuries, used to govern the apportionment of sacrifices and expenditures made to protect property in a common maritime adventure. Source: https://charles-taylor-group.s3.amazonaws.com/production/filer_public/e8/c6/e8c600ba-e591-43e3-a9d0-3a885e61b776/rhl_-_york_antwerp_rules_2016_-_a_summary_of_the_changes.pdf
“Montreal Convention”, which aims to unify and modernize the legal rules governing the liability for the carriage of passengers, baggage, and cargo in international air transportation. Source: https://store.icao.int/en/convention-for-the-unification-of-certain-rules-for-international-carriage-by-air-doc-9740

**dressing emerging challenges such as algorithmic transparency, cross-border data flows, model alignment, and system safety, thereby innovatively constructing a global AI safety and governance system that balances technological features with public interests.**

Table 1: Lessons from Global Multilateral Governance Frameworks in Other Sectors

| Governance Domain | Representative Governance Bodies | Typical Governance Measures | Core Mechanisms and Frameworks | Lessons for Global AI Governance |
|---|---|---|---|---|
| Nuclear Safety | International Atomic Energy Agency (IAEA) | Prevents nuclear proliferation through a combination of rigorous verification mechanisms and technical assistance; ensures safe operation of nuclear facilities and nuclear accident information notification | · *Treaty on the Non-Proliferation of Nuclear Weapons (NPT)*<br>· *Convention on Nuclear Safety*<br>· *Convention on Early Notification of a Nuclear Accident*<br>· *Convention on Assistance in the Case of a Nuclear Accident or Radiological Emergency* | Comprehensive governance that is both coercive and cooperative, and promotes transnational trust and collaboration |
| Climate Change | Intergovernmental Panel on Climate Change (IPCC) | Examines projections of global warming trends, guides the formulation of national emissions reduction policies, and promotes transnational collaboration on climate action through multiple rounds of interdisciplinary scientific assessments, expert collaboration and data-sharing platforms | · *United Nations Framework Convention on Climate Change (UNFCCC)*<br>· *Kyoto Protocol*<br>· *Paris Agreement* | Strengthening forward-looking and scientific support through science-supported policies and a dynamically updated governance model |
| International Trade | World Trade Organization (WTO) | Promotes a balance of rights and interests through a systematic dispute settlement mechanism, arbitration decisions, and negotiations on tariff disputes, intellectual property rights conflicts, and trade barriers | · *Agreement Establishing the World Trade Organization*<br>· *General Agreement on Tariffs and Trade (GATT 1994)*<br>· *General Agreement on Trade in Services (GATS)*<br>· *Agreement on Trade-Related* | Efficient dispute resolution and arbitration mechanisms with clear rules and processes that take into account the interests of all stakeholders |
| Public Health | World Health Organization (WHO) | Prevents global outbreaks through outbreak information reporting, trans-regional source tracking, emergency quarantine and resource allocation mechanisms | · *WHO, The International Health Regulations (IHR)*<br>· *WHO Framework Convention on Tobacco Control* | Governance model of cross-border linkage and rapid response to strengthen global crisis prevention, control and emergency response collaboration |
| Global Aviation & Maritime Transport | International Civil Aviation Organization (ICAO), International Maritime Organization (IMO) | Ensures the safe operation of aviation and shipping and promotes the interoperability of cross-border rules through the formulation of technical standards, authoritative investigations of safety incidents, and assessments of airworthiness and ship safety | · *International Convention for the Safety of Life at Sea (SOLAS)*<br>· *United Nations Convention on the Law of the Sea (UNCLOS)*<br>· *York-Antwerp Rules*<br>· *Convention on International Civil Aviation (Chicago Convention)*<br>· *Montreal Convention* | Governance model of synergistic standards and continuous oversight to promote harmonization of standards and cross-system risk management |

— PART —

02

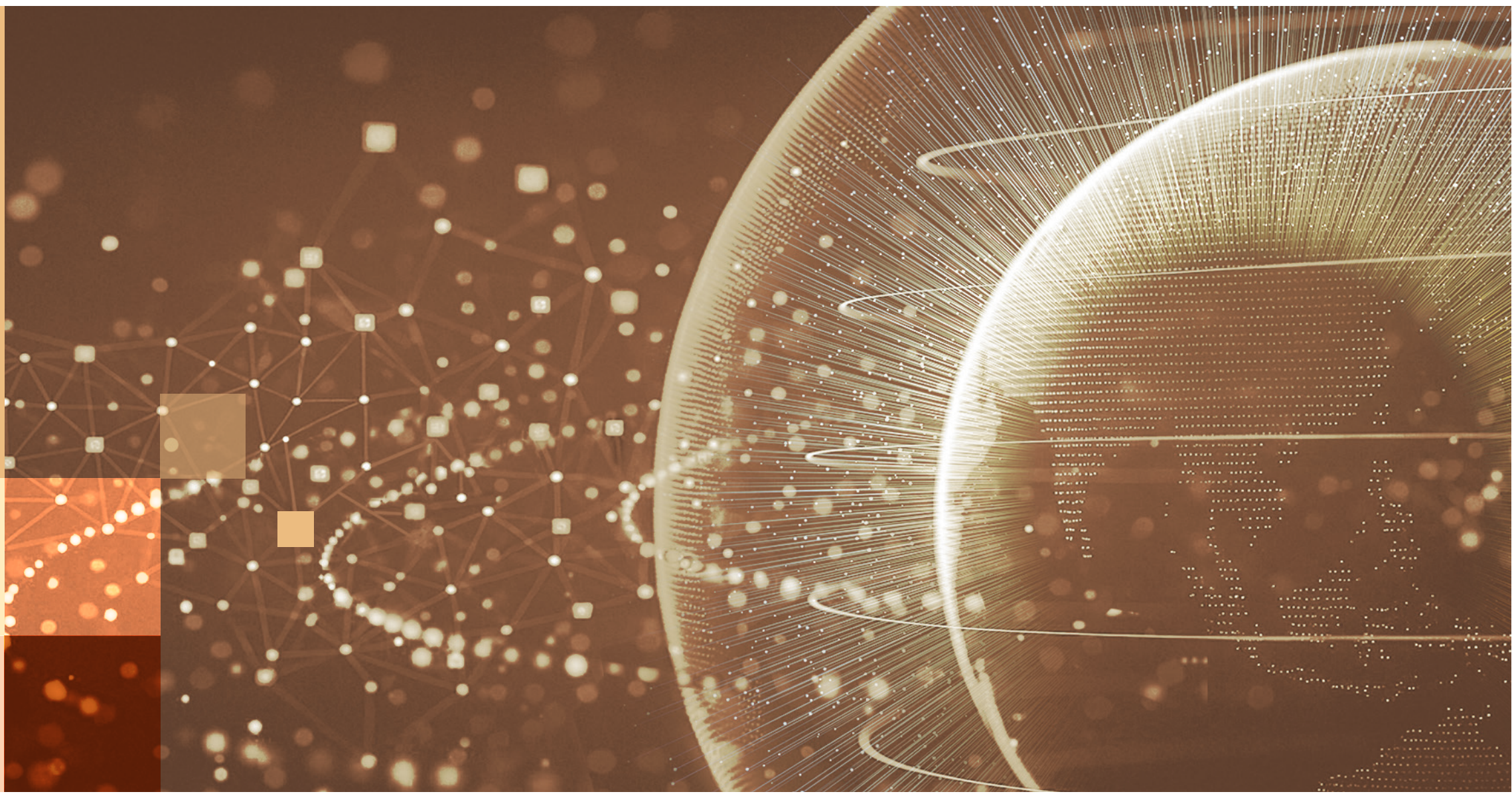

# Ensuring Safety : Responding to the Rapid Transformation and Major Risks of AI

This chapter analyzes the multi-layered risks brought about by the development of AI from a safety perspective and explores the design of corresponding global AI safety and governance mechanisms. The first two sections focus respectively on the uncertainties arising from rapid technological iteration and the risks of abuse and malicious use in the context of widespread application, revealing challenges such as governance lag, insufficient tools, and cross-border regulatory complexities. The third section then addresses systemic major risks that could affect global strategic security, critical infrastructure, and human survival, emphasizing the urgent need for forward-looking assessment, international coordination, and emergency intervention mechanisms to build a dynamically adaptive and transnationally collaborative AI safety and governance system.

## (a) Addressing the Risks from Rapid Iteration and Uncertainty of AI Technologies

Currently, the development of AI technology is characterized by **fast iteration speed, unpredictable breakthrough paths, and complex model capabilities and safety properties. First,** the level of technological capability iterates rapidly, with its rate of improvement significantly surpassing traditional linear predictions. For example, the parameter scale and reasoning performance of large language models have achieved orders-of-magnitude breakthroughs in a short time[88], transitioning from basic dialogue to demonstrating high-level reasoning abilities in fields like mathematics and programming in less than three years. Simultaneously, new technologies continue to emerge, such as diffusion models and self-supervised learning par-

88. Within a year, cutting-edge AI models made significant breakthroughs in a number of demanding benchmark tests. Source: https://hai.stanford.edu/assets/files/hai_ai_index_report_2025.pdf

adigms being rapidly applied to multiple types of tasks; multimodal large models continue to make breakthroughs; and technologies like reasoning AI, agentic AI, and embodied intelligence are rapidly maturing. **Second,** technological path breakthroughs are highly unpredictable. For instance, the Transformer architecture replaced Recurrent Neural Networks as the mainstream paradigm for natural language processing in a short period; some advanced capabilities do not improve linearly with model scale but instead emerge[89] suddenly after parameters or computation reach a specific critical point. **Finally,** the technological system possesses inherent vulnerabilities and lack of robustness, while understanding of its complex safety properties remains insufficient. For example, large models are prone to uncontrollable erroneous outputs under adversarial attacks and are still difficult to defend effectively; current understanding of the mechanisms behind model hallucination, bias, and safety alignment is also still inadequate.

**The aforementioned characteristics of AI technology development further pose multiple challenges to traditional governance mechanisms. First, direction is hard to predict, making proactive deployment difficult.** The timing and path of critical technological breakthroughs are highly uncertain, making it difficult for governance measures to conduct effective forward-looking deployment, often leading to reactive responses; simultaneously, the lack of systemic risk monitoring indicators and capabilities also causes regulatory measures to often lag behind the technological reality. **Second, the pace is hard to keep up with, and dynamic adjustment is insufficient.** Global multilateral negotiations and international rule-making typically operate on timescales of years, creating a significant speed gap with AI's monthly or even weekly iteration cycles; the formulation and updating of governance rules often lag behind rapid technological evolution; regulatory measures across different countries and regions are also noticeably out of sync, creating potential governance gaps. **Third, tools are imperfect, and technical support is limited.** As regulatory measures lack systematic risk assessment and intervention tools, particularly since monitoring indicators and assessment platforms for model application safety risks are not yet mature, the comprehensiveness and effectiveness of regulatory agencies in risk identification, assessment, and intervention are constrained.

**In response, building a global safety and governance system adapted to AI technology development should adhere to the principles of being scientific, agile, and collaborative, focusing on strengthening the following mechanisms. First, establish collaborative mechanisms for technology tracking and risk early warning.** Within multilateral frameworks like the United Nations, promote the formation of a transnational technology monitoring network, facilitating proactive disclosure and sharing of information on cutting-edge AI model capabilities and safety incidents among stakeholders. Regularly organize and publish international joint assessment tests of cutting-edge model capabilities and risks, collectively enhancing shared scientific understanding and early warning capabilities for emerging and potential risks. **Second, establish dynamic updating and mutual recognition**

89. Wei J, Tay Y, Bommasani R, et al. Emergent abilities of large language models[J]. arXiv preprint arXiv:2206.07682, 2022.

**mechanisms for governance rules.** Promote relevant international standards organizations to establish and improve regular, periodic review and update procedures for AI safety standards. Encourage major economies to take the lead in achieving mutual recognition of test results and certifications in high-risk application areas. Through dynamic rule correction and interoperability, effectively prevent regulatory arbitrage and ensure agile and coordinated governance. **Third, jointly build an ecosystem of common safety evaluation tools and platforms.** Encourage, through international cooperation, the joint development of open, highly interoperable AI safety testing platforms, benchmark datasets, and risk assessment tool libraries. Build a globally shared technical toolbox to provide precise, effective, and coordinated capability support for national regulatory agencies, continuously enhancing regulatory effectiveness.

### (b) Addressing the Risks of the Broad Application, Abuse, and Malicious Use of AI

**In recent years, a new generation of AI technologies, represented by large-scale generative models and the open-source ecosystem, has significantly lowered development and application barriers, driving the rapid proliferation of capabilities.** For example, ChatGPT reached hundreds of millions of monthly active users within just two months of its launch in late 2022[90], while open-source models like DeepSeek have further promoted capability dissemination and cost reduction. **Against this backdrop, AI capabilities are rapidly embedding into various aspects of social production, public services, scientific research and innovation, and information dissemination, with the technology's influence rapidly expanding to all sectors of society. In the socio-economic domain,** AI is deeply embedded in financial trading, smart manufacturing, and supply chain management, significantly enhancing resource allocation efficiency and productivity levels, but it also brings risks such as market concentration, employment structure adjustments, and technological dependency, posing challenges to economic security and fair competition. **In the social livelihood domain,** AI is widely used in education, healthcare, and public services, improving the accessibility and precision of services, while issues such as algorithmic decision-making opacity, data bias, and the digital divide have raised ongoing concerns about fairness and privacy protection. **In the scientific and technological innovation domain,** AI aids in protein structure prediction, new material discovery, and drug development, accelerating scientific progress, but problems regarding research ethics, ownership of results, and data verifiability are becoming increasingly prominent, posing new tests for research integrity. **In the legal and institutional aspects,** generative AI has profoundly impacted content creation and dissemination mechanisms, leading to legal application disputes over areas like intellectual property attribution, infringement liability determination, and the scope of fair use, posing new requirements for the adaptability of the current legal system. **In the ethical and social values domain,** AI-generated content may blur the boundary between real and fictional, while risks of model bias, manipulation, and emotional in-

90. ChatGPT records fastest growing user base. Source: https://www.reuters.com/technology/chatgpt-sets-record-fastest-growing-user-base-analyst-note-2023-02-01/

tervention also pose new ethical challenges to human-machine relationships and social value systems.

**At the same time, the abuse and malicious use of AI are continuously amplifying social risks, increasing the complexity and urgency of social governance.** AI-assisted fraudulent activities are becoming more sophisticated and frequent; data shows that in 2024, a deepfake attack occurred on average every five minutes, and AI-assisted digital document forgery increased by 244% year-on-year, surpassing physical forgery as the primary fraud method for the first time[91]. Simultaneously, automated applications in areas like recruitment and credit approval have triggered social discrimination[92]. Fabricating statements by political figures and creating false public opinion situations have repeatedly interfered with elections and diplomatic activities[93]. User profiling and psychological manipulation based on cross-platform behavioral data have made privacy violations more concealed and scaled due to AI's powerful data correlation and analysis capabilities[94]. These risks not only have apparent social harms but may also erode the foundation of social trust.

**From a global governance perspective, the governance measures of any single country are already insufficient to fully address the widespread social impact and risks of AI, while an effective global coordination framework still faces multiple challenges. First, there is insufficient consensus on risk perception.** Countries have certain differences in AI application scenarios and standards for risk assessment, classification, and grading, leading to varying regulatory demands and methods, which constrain the formation of a synergistic governance force. **Second, cross-border risk prevention mechanisms have yet to be fully developed.** The cross-border dissemination of AI-generated content lacks internationally recognized identification and certification standards, and insufficient traceability capabilities increase the difficulty of governing risks like disinformation. **Third, there are obstacles to cross-border law enforcement cooperation.** There is a lack of standardized procedures for cross-border investigations, evidence preservation, and judicial cooperation targeting AI abuse, affecting actual governance effectiveness.

**Therefore, there is an urgent need to establish and improve corresponding mechanisms in several aspects, so as to balance AI development with governance and build a safer, more orderly intelligent society. First, promote the formation of an internationally consensual risk governance framework.** Through multilateral consultation mechanisms, promote the establishment, improvement, and coordinated mutual recognition of AI risk classification and grading standards among countries, gradually building an internationally consensual frame-

91. Deepfake attacks occur once every five minutes. Source: https://www.helpnetsecurity.com/2024/11/22/ai-assisted-fraud-rise/
92. Artificial intelligence program developed by Amazon.com to screen resumes found to be sexist. Source:https://paper.people.com.cn/zgcsb/html/2023-09/18/content_26017767.htm
93. Artificial intelligence-generated content is posing a threat to the election process in a number of ways, and there is an urgent need for a multi-pronged response. Source:https://www.iiss.pku.edu.cn/__local/3/7F/3D/9EA4A3B30950C7A22DAC5E56962_861E8C97_60B97. pdf.
94. Algorithms in artificial intelligence are often used for user profiling to track, summarize, and analyze massive amounts of information to extract private information about users, which can lead to commercial abuses such as big data ripening and discriminatory pricing. Source: https://www.zhonglun.com/research/articles/8670.html

work for AI risk classification and grading, laying the foundation for precise and coordinated cross-border risk identification and regulation. Second, establish a collaborative system for cross-border content governance. Relying on existing relevant international standards organizations, promote the coordination of international standards for traceability technologies like digital watermarking and content authentication, gradually building a global AI content traceability and authentication network, and establish a globally shared traceability database to enhance the cross-border identifiability and traceability of AI-generated content. Encourage multi-stakeholder collaborative participation in the design and implementation of relevant mechanisms to collectively improve the capacity to identify, track, and respond to disinformation. In this regard, China's "Measures for Labeling of AI-Generated Synthetic Content"[95], which took effect in September 2025, along with its supporting mandatory national standard[96], provide a reference for standardizing the governance of AI-generated content and lay a practical foundation for establishing a cross-border collaborative system for content governance. **Third, improve cross-border law enforcement cooperation mechanisms.** Within the framework of international organizations, build an AI safety and security threat information-sharing network, promote the standardization of procedures for cross-border investigations and electronic evidence preservation, and gradually establish monitoring and coordinated disposal mechanisms targeting the abuse and malicious use of AI technology, enhancing the global capacity to respond to and address cross-border illegal activities.

## (c) Anticipating and Preventing Major Global Risks of AI

**Beyond the risks arising from rapid technological iteration and widespread application, abuse, and malicious use, AI may also pose global and systemic risks, including potential threats to strategic security, international peace, and even human survival.** These risks are characterized by their transnational, systemic, and strategic nature. Their impact scope may cover global critical infrastructure such as energy, transportation, finance, and communications. They are not confined to a single country or localized application but may cross national borders, industrial systems and social structures[97] [98], posing potential major threats to global security and human well-being. **At the same time, as AI technologies continue to advance toward Artificial General Intelligence and potential Superintelligence, the systemic risks such developments could trigger are becoming increasingly evident.** AI possessing the ability to integrate resources and execute complex tasks across domains will cause risks to expand from local to global. For example, in civilian scenarios, it may trigger transnational public crises such as energy supply disruption, transportation system paralysis, and global financial system failure; in military and security domains, the loss of con-

95. China's "Measures for Labeling of AI-Generated Synthetic Content" takes effect on September 1, 2025. Source: https://www.gov.cn/zhengce/zhengceku/202503/content_7014286.htm
96. "Cybersecurity technology—Labeling method for content generated by artificial intelligence". Source: https://openstd.samr.gov.cn/bzgk/std/newGbInfo?hcno=F32EA2A561F1886CD8D606513512D547
97. Goldman Sachs Report Predicts AI Could Impact 300 Million Full-Time Jobs. Source: https://www.goldmansachs.com/insights/articles/generative-ai-could-raise-global-gdp-by-7-percent
98. A widespread IT outage caused by a CrowdStrike update. Source: https://www.cisa.gov/news-events/alerts/2024/07/19/widespread-it-outage-due-crowdstrike-update

trol of highly autonomous lethal weapon systems may trigger strategic defense miscalculations[99], rapid escalation of local conflicts, and even touch upon nuclear safety[100]. Furthermore, the global concentration of AI capabilities and the cross-border flow of technology mean that a regulatory gap in any single country or region could escalate into a global systemic risk—a manifestation of the "barrel effect". This reality underscores the critical need for internationally coordinated governance and collective risk mitigation.

**At the global mechanism level, there have been some forward-looking efforts to address the aforementioned risks.** For instance, in the area of AI arms control and strategic security, countries continue discussions on Lethal Autonomous Weapons Systems (LAWS) within the framework of "The Convention on Certain Conventional Weapons (CCW)", and it is widely recognized that human control is essential to ensure responsibility and accountability, compliance with international law, and ethical decision-making[101]; The Global Commission on Responsible Artificial Intelligence in the Military Domain (GC-REAIM) released its report, which identifies that the decision to authorize the use of nuclear weapons should remain under human control as one of its core recommendations[102]; Meanwhile, the leaders of China and the US reached a consensus on maintaining human control over nuclear weapon use[103]; The "International AI Safety Report" [104]promotes the international scientific community to jointly assess the capabilities and potential risks of AGI and related frontier technologies. In cross-domain safety testing, joint experiments and the development of standard tools have been preliminarily applied in risk management areas including AI and Chemical, Biological, Radiological, and Nuclear risks (CBRN), such as establishing shared assessment platforms[105], simulated joint tests, and cross-border capability verification.

**Nonetheless, current global mechanisms remain insufficient in addressing potential major risks. First, participation in international joint assessments of frontier risks is limited.** Unified standards and joint testing processes covering transnational security-sensitive scenarios like CBRN are currently only being explored among a limited number of countries. **Second, institutional constraints on the sources of major risks are insufficient.** Technology flow and regulatory differences may lead to the cross-border abuse of high-risk technologies, making global systemic risks difficult to control effectively. **Third, emergency intervention and joint response capabilities are limited,** lacking rapidly activatable international mechanisms to halt dangerous

99. Lethal autonomous weapons systems pose a major challenge to international security and humanitarianism. Source: https://www.siis.org.cn/updates/cms/old/UploadFiles/file/20200307/202002006%20%E9%BE%99%20%20%E5%9D%A4.pdf
100. AI could spark a "third revolution in warfare". Source: https://safe.ai/ai-risk
101. United Nations, "Lethal Autonomous Weapons Systems: Report of the Secretary-General". Source: https://docs.un.org/en/A/79/88
102. The Report by the Global Commission on Responsible Artificial Intelligence in the Military Domain (GC REAIM) proposes five core recommendations including "Agree, at a legally binding level, that the decision to authorize the use of nuclear weapons should remain under human control". Source: https://hcss.nl/wp-content/uploads/2025/09/GC-REAIM-Strategic-Guidance-Report-Final-WEB.pdf
103. The two heads of state confirmed that the decision to maintain human control over the use of nuclear weapons should be upheld. Source: https://www.gov.cn/lianbo/bumen/202411/content_6987686.htm
104. Y. Bengio et al. , "International AI Safety Report (DSIT 2025/001)". Source: https://www.gov.uk/government/publications/international-ai-safety-report-2025
105. The International Network of AI Safety Institutes has launched a joint testing exercise targeting autonomous agent systems based on large language models. Source: https://www.aisi.gov.uk/blog/international-joint-testing-exercise-agentic-testing

R&D or block potentially malicious deployments.

**In response, it is necessary to establish forward-looking, systematic, collaborative, and agile mechanisms to further enhance international capacities for preventing and managing major global AI risks. First, a consensus should be formed on jointly addressing the risks of AI technology running out of human control.** This includes promoting fundamental principles of trustworthy AI that encompass technological, ethical, and managerial dimensions; setting relevant requirements for the use of AI in sensitive domains such as nuclear, biological, chemical, and missile applications; strengthening traceability management of AI's ultimate uses; and fostering international agreement to prevent the misuse and abuse of AI technologies. These efforts aim to guard against uncontrollable risks that could threaten human survival and development, ensuring that AI evolves safely, reliably, and controllably, and remains consistently under human oversight. Second, a globally participatory mechanism should be established for monitoring frontier AI models and assessing their capabilities. This mechanism should focus on high-risk applications and technological domains, leveraging cross-border joint testing platforms, standardized risk indicators, misuse monitoring, and regular capability assessments to ensure that potential major risks are widely identified and reported. Third, an international governance system should be developed for critical resources. By focusing on highly sensitive areas such as computing power, data, and open-source algorithms, efforts should be made to establish transparent international rules and coordinated governance frameworks, thereby reducing the spread of technology misuse risks at their source. Fourth, emergency intervention and joint response mechanisms should be put in place. When potential catastrophic or strategic risks are identified, these mechanisms should enable the rapid initiation of cross-national coordinated measures—such as suspending hazardous R&D, restricting financial flows and technology diffusion, or halting dangerous deployments—to maintain control over critical systems and ensure that AI applications always serve the well-being of humanity.

— PART —

03

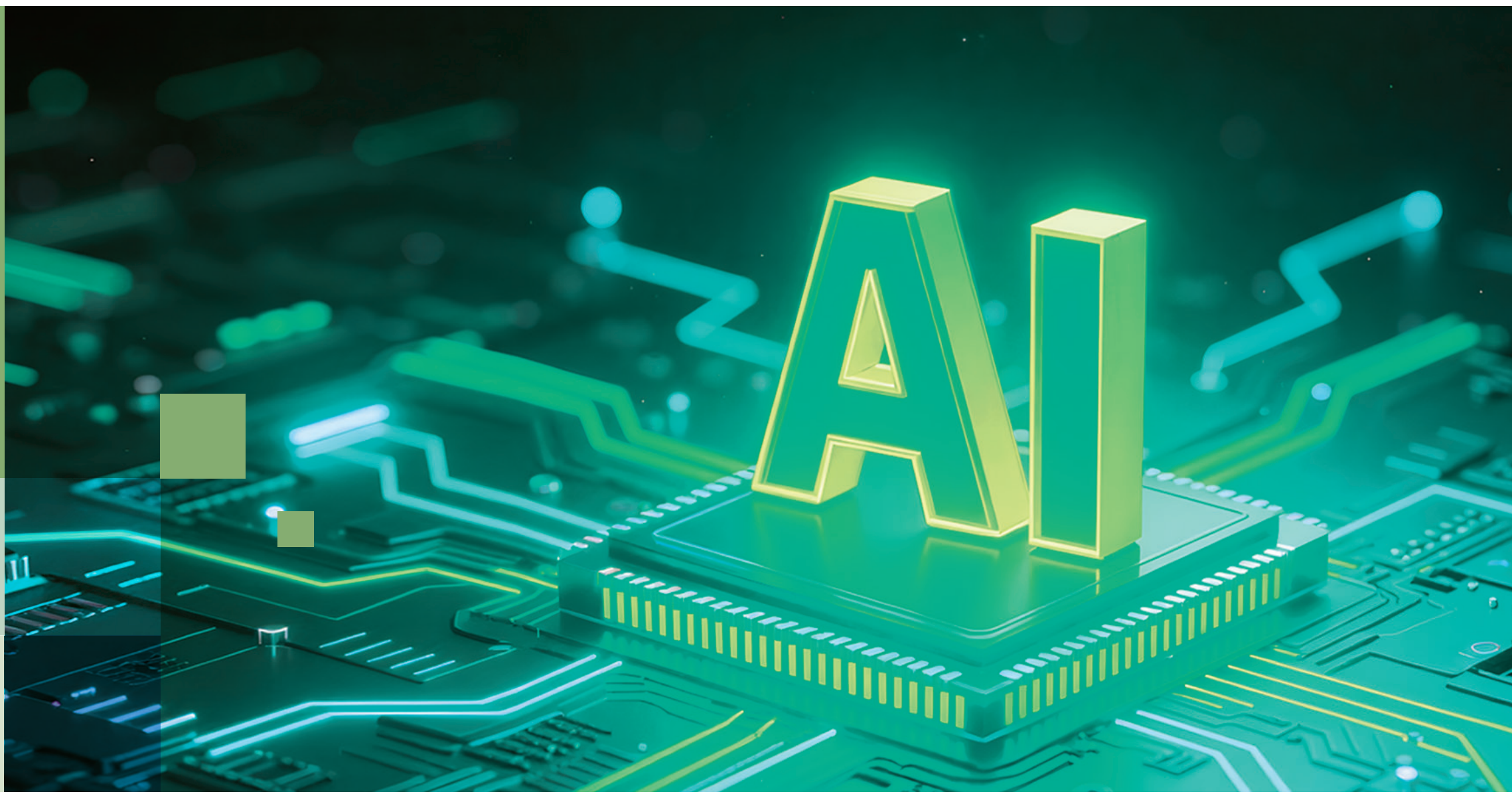

# Ensuring Inclusiveness: Balancing Global Development and Governance Demands in AI

While AI presents unprecedented opportunities for global development, it also highlights disparities among nations in technological capacity, resource endowment, and institutional preparedness. These differences not only lead to diverse development demands but also complicate the coordination of global AI governance. To achieve inclusive governance, the international community must strike a balance between respecting diversity and promoting common interests. This chapter focuses on "ensuring inclusivity": first, it analyzes the structural divides in global AI development and the resulting diverse demands; second, it explores pathways to safeguard the rights of all countries to equally develop and utilize AI; finally, it proposes institutional arrangements to achieve representativeness and inclusivity within global governance mechanisms, forming a fair, inclusive, and sustainable governance system and cooperative framework.

## (a) Recognizing the Global Development Divide and Diverging Governance Priorities

**The global development of AI exhibits significant structural imbalances, with resources, capabilities, and institutional advantages heavily concentrated in a few developed economies and large technology corporations.** The "AI Index Report 2025" [106]reveals that in 2024, private sector AI investment in the United States reached approximately $10.91 billion, nearly 12 times that of China ($930 million). Meanwhile, the "Technology and Innovation Report 2025" [107]highlights that about 40% of global corporate research and development expenditure is concentrated in 100 large technology firms, most of which are headquartered in the United States, the Europe-

106. Stanford Institute for Human-Centered AI (Stanford HAI), "AI Index Report 2025". Source: https://hai.stanford.edu/ai-index/2025-ai-index-report
107. United Nations Conference on Trade and Development (UNCTAD), "Technology and Innovation Report 2025". Source: https://unctad.org/system/files/official-document/tir2025_en.pdf

an Union, or Japan. This high concentration of capital and technology enables a small number of countries to hold a dominant position in algorithmic innovation, standard-setting, and industrial ecosystems.

**Disparities in institutional and governance capabilities are equally pronounced.** The "Artificial Intelligence Preparedness Index (AIPI)" published by the International Monetary Fund[108] reveals that high-income countries have significantly better AI preparedness than emerging market and low-income countries. The "Government AI Readiness Index 2024" report[109] further indicates that developed countries hold significant advantages in the "technology sector pillar" dimension, whereas low- and middle-income countries generally lag in "government governance" and "data infrastructure". A World Bank report[110] also highlights that talent shortages, inadequate information and communication technology infrastructure, and dependence on core technologies constitute major bottlenecks for the adoption and governance of AI in developing countries.

**The development divide has further led to significant disparities in governance demands.** Technologically advanced countries tend to focus more on frontier breakthroughs, competitive advantages, and security risk control, while developing countries prioritize technology accessibility, capacity building, and localized applications, aiming to leverage AI to drive economic growth, improve public services, and bridge the digital divide. These differing structural conditions shape countries' diverse positions in AI policy objectives, risk perceptions, and international rule-making negotiations. **At the same time, such differences further exacerbate the complexity and coordination challenges of global governance. First, building international consensus remains difficult.** Divergences in technological development levels, economic interests, and risk perceptions among nations lead to fragmented positions on key governance issues, resulting in high coordination costs. **Second, the advancement of a unified regulatory framework faces obstacles.** Differing interpretations of safety thresholds, regulatory principles, and ethical standards undermine the foundation for mutual recognition of transnational rules and institutional synergy, thereby constraining overall governance effectiveness. **Third, there is a notable imbalance in governance participation.** Technologically leading countries dominate standard-setting and agenda formation, while many developing countries lack meaningful engagement. This not only undermines the fairness and representativeness of governance but could also intensify future fragmentation and conflict within the global governance architecture.

**To achieve universality and inclusiveness in global governance, it is essential to accommodate the development stages and interest of different countries.** To this end, the international community could focus on the following three dimensions: **First, building open and inclusive multilateral consultation platforms** to ensure countries at different development stag-

108. International Monetary Fund (IMF), "AI Preparedness Index (AIPI)". Source: https://www.imf.org/en/Blogs/Articles/2024/06/25/mapping-the-worlds-readiness-for-artificial-intelligence-shows-prospects-diverge
109. Oxford Insights, "Government AI Readiness Index 2024". Source: https://oxfordinsights.com/ai-readiness/ai-readiness-index/
110. World Bank, "Bridging the AI Divide". Source: https://openknowledge.worldbank.org/bitstreams/82fbc048-b723-4818-bb91-9a4c8855daf1/download

es and with diverse institutional backgrounds can participate equally in setting governance agendas, formulating rules, and negotiating technical standards, thereby enhancing the legitimacy and inclusiveness of global governance. **Second, establishing practical and effective international cooperation mechanisms** that, while respecting national sovereignty and "AI governance rights", provide technical assistance, experience sharing, and resource coordination to support developing countries in enhancing their AI governance and innovation capabilities. This would promote technological inclusivity, co-development of infrastructure, and talent collaboration, bridge the digital and AI divide, and continuously foster the formation of a consensus-based international cooperation framework. **Third, implementing fair and transparent decision-making procedures** by embedding open, traceable, and accountable operational rules within governance mechanisms. This would minimize information asymmetry, strengthen international mutual trust, and lay an institutional foundation for continuously addressing global risks.

## (b) Ensuring the Equality of Development and Application of AI for All Countries

An inclusive and equitable global AI safety and governance framework must respect the diversity of national development models and cultural contexts, foster collaborative governance and technological inclusivity, and safeguard the rights of all countries to equally develop and utilize AI. **However, in reality, deep-seated inequalities persist in global digital development rights, with the AI divide continuing to widen, constraining equal participation and shared benefits in the AI domain across nations.**

**The root causes are primarily reflected in three aspects. First, technologically advanced countries continue to consolidate their structural advantages in core resources.** Their dominance in foundational resources such as computing power, algorithms, and data continues to strengthen, while the international circulation of high-end chips and high-performance computing infrastructure faces numerous restrictions. For example, Global North countries possess 75% of the world's most powerful supercomputers, while the entire African continent accounts for less than 1% of these high-performance systems. This scarcity imposes significant cost pressures on African countries in developing AI—relative to their per capita economic levels, the cost of acquiring core AI computing equipment such as Graphics Processing Units is often 10 to 30 times higher for African nations compared to developed countries[111]. **Second, the international community still lacks a systematic mechanism for AI development assistance and resource coordination.** Within existing global innovation cooperation frameworks, no dedicated, long-term, and institutionalized support channels for AI have been established. Developing countries often struggle to overcome the constraints of high research and development investment and infrastructure construction costs, resulting in limited technological self-sufficiency. **Finally, the fragmented and short-term nature of capacity-building cooperation also hinders sus-**

111. African countries face high costs in acquiring core AI computing equipment such as graphics processing units. Source: https://www.aihubfordevelopment.org/green-compute-coalition

**tainable development.** Most current projects primarily consist of phased assistance, lacking continuous institutional support and knowledge transfer mechanisms. This leads to a persistent lag in talent cultivation, institutional development, and risk governance capabilities in less developed regions, failing to keep pace with the speed of technological iteration.

**The global imbalance in AI development calls for an international governance system with greater inclusivity and coordination. Specifically, three measures should be prioritized: First, establish a controllable and secure multilateral mechanism for technology sharing and transfer.** By facilitating the orderly flow of key elements such as algorithms, computing power, and data, support technologically latecomer countries in achieving capability leapfrogging. **Second, innovate financing models.** Leverage multilateral institutions like the World Bank and regional development banks to set up special AI development funds, issue AI sustainable development bonds, link debt repayment to technological capacity-building outcomes (for example, by providing debt relief to countries that meet preset development targets), and establish Public-Private Partnership (PPP) financing pools to attract private sector participation in the construction of computing power centers in developing countries[112]. **Third, develop a systematic and long-term capacity-building cooperation mechanism,** establishing an international support network for talent cultivation, regulatory frameworks, standard alignment, and risk governance to assist countries in enhancing local governance and adaptive development capacities.

### (c) Ensuring Representation and Inclusiveness in Global AI Safety and Governance Mechanisms

**A fair and credible international governance mechanism should ensure the representation and inclusiveness of countries in decision-making processes.** Representativeness requires that the governance structure fully incorporates countries at different development stages and with diverse interests, guaranteeing them substantive participation in key processes such as international agenda-setting, rule negotiation, and standard formulation. Inclusiveness, on the other hand, emphasizes that through equal dialogue and extensive consultation, the governance system achieves effective coordination while respecting the diversity of values and development paths.

**Currently, decision-making power and discourse power in global AI governance remain highly concentrated in a small number of countries.** According to statistics[113], among the 193 UN Member States, only 7 countries have participated in the seven major AI governance initiatives proposed in recent years, while 118 Member States were completely absent—primarily countries from the Global South. The “Global Index for AI Safety 2025”[114], which surveyed 40 countries, similarly revealed that only 6 countries have signed all five key international AI safety declarations and initiatives in recent years, highlighting a severe lack of participation

112. United Nations, “Innovative voluntary financing options for artificial intelligence capacity-building”. Source: https://digitallibrary.un.org/record/4085951?ln=en&v=pdf#files
113. United Nations, “Governing AI for Humanity” Final Report. Source: https://www.un.org/sites/un2.un.org/files/governing_ai_for_humanity_final_report_en.pdf
114. “Global Index for AI Safety (GIAIS) 2025”. Source: https://agile-index.ai/global-index-for-ai-safety

and representativeness in global safety and governance mechanisms.

**The reasons for this situation primarily include three points: First,** there is a lack of binding credibility and commitment enforcement mechanisms at the international level. Some major countries can flexibly dominate or even withdraw from key agendas by virtue of their advantages, undermining the overall effectiveness and authority of multilateral rules. **Second,** as global common baselines and redlines for safety and ethics that all countries abide by have not yet been established, there are significant differences in regulatory thresholds among countries. Developed countries often dominate the discourse by taking the lead in setting standards and establishing evaluation systems, which virtually raises the compliance and participation costs for late-developing countries. **Third,** the principle of differentiated responsibility is not reflected in the governance structure. Developing countries often passively bear heavy compliance burdens but hardly have an impact on substantive decision-making, resulting in their demands failing to be effectively incorporated into international rules.

**To build a fairer, more inclusive, and binding global AI governance system, efforts should focus on developing the following mechanisms: First, an international dynamic assessment and reputation-linked mechanism for compliance should be established,** so as to conduct continuous evaluations of countries' participation, link the assessment results to their international reputation, and curb arbitrary withdrawal behaviors as much as possible. **Second, efforts should be made to promote the establishment of a compliance hearing mechanism for the decision-making process of global AI governance.** Centering on core principles such as representativeness, transparency, procedural legitimacy, and consensus-building, this mechanism will conduct multilateral review and supervision over the formulation process of major decisions, so as to prevent legitimacy deficits and governance trust crises caused by procedural injustice. **Third, a differentiated responsibility-sharing mechanism based on each country's development level, technological capacity, and governance foundation should be implemented.** While ensuring that technologically leading developers bear the responsibility for risk assessment and governance of frontier models, it is equally crucial to safeguard the effective exercise of supervisory rights by affected parties, thereby promoting fairness, effectiveness, and sustainability in global AI governance.

— PART —

# 04

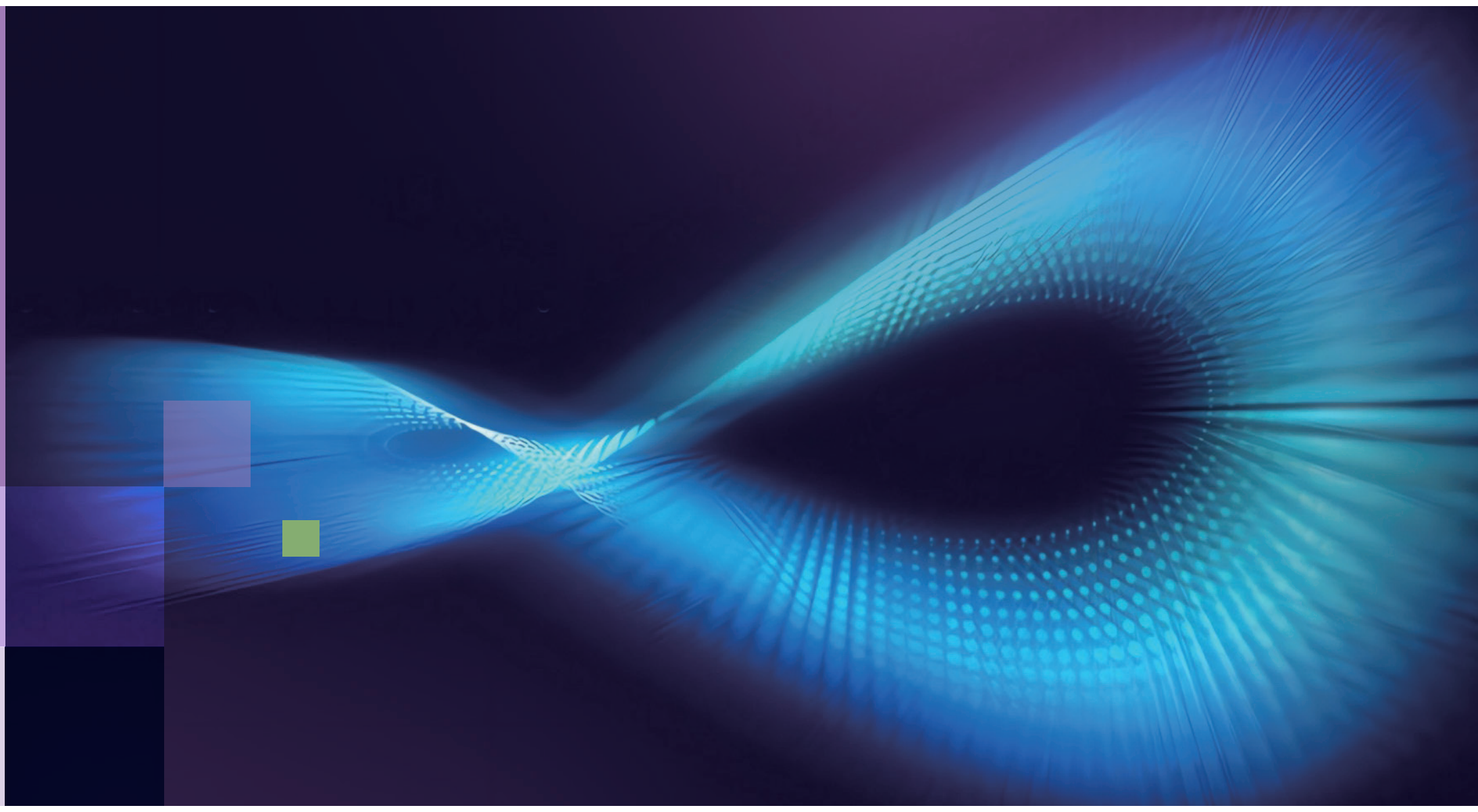

## Clarifying Responsibilities: Promoting Coordinated and Effective Multi-Stakeholder Action

This chapter analyzes the complex interactions and challenges among multiple stakeholders in AI safety and governance, identifies the differences and complementarities among various parties in their role positioning, resource endowments, and value orientations, and emphasizes the limitations of a single actor dominating rule-making and the potential impacts this may have on the legitimacy, effectiveness, and sustainability of governance. It further clarifies the differentiated roles and responsibility boundaries of core entities such as national governments, international organizations, enterprises, research institutions, and the public, and proposes the establishment of a systematic coordination mechanism centered on compliance review, risk monitoring, dispute mediation, and standard mutual recognition. This mechanism aims to address issues such as ambiguous responsibilities, fragmented rules, and implementation difficulties, and promote the formation of a global AI safety and governance system focused on clarifying roles and responsibilities and strengthening effective coordination among all stakeholders.

### (a) Complex Interactions and Challenges Among Multiple AI Stakeholders

**The inherent openness of AI technologies, along with their cross-domain applicability, makes the broad participation of multiple stakeholders a necessary condition for their governance.** Governments, international organizations, enterprises, research institutions, and the public differ in their responsibilities, expertise, resource endowments, and value orientations, while also complementing one another. For example, governments hold policy-making and regulatory resources; enterprises possess innovation capacity and technical expertise; research institutions provide scientific assessment and knowledge support; international organizations promote collaboration and standard alignment; while the public plays a crucial role in shaping value orientation and exercising

social oversight. A governance framework that relies solely on a single actor or dimension in rule-making would not only undermine the legitimacy of governance arrangements and public trust, but also constrain its effectiveness in addressing complex risks and ensuring long-term sustainability.

**The complex interactions among multiple stakeholders have significantly increased the difficulty of coordination in AI safety and governance.** Differences in roles, resource endowments, and goal priorities make stakeholders both interdependent and mutually constraining: states rely on enterprises' technological capabilities and research institutions' knowledge contributions for policy-making; international organizations depend on member states' political commitments and civil society's value advocacy to advance agendas; enterprises' development hinges on national policy environments and public trust; and scientific innovation benefits from corporate investment while remaining subject to social oversight.

**In the global AI safety and governance framework, the lack of effective participation and coordination among multiple stakeholders may give rise to multiple risks. At the macro level,** competition among countries for discursive dominance in the AI field could further fragment the technological ecosystem and governance rules, raise technological barriers, and even foster a divergence of governance mechanisms based on differing values or security concerns. **At the practical level, First,** the agenda-setting of global governance may lose its due representativeness and balance, undermining the neutrality and credibility of multilateral mechanisms. **Second,** public policy-making may be overly influenced by large tech enterprises, leading to regulatory leniency or even "regulatory capture", which may undermine the adequate protection of the public interest. **Third,** the independence and openness of scientific research face increasing pressure from commercialization, with the open-source ecosystem correspondingly constrained. **Fourth,** enterprises' pursuit of technological monopolies and commercial profit may come into conflict with public demands for fairness, accountability, and rights protection, potentially exacerbating social divisions. **Therefore, ensuring sufficient participation and effective coordination among multiple stakeholders within the global AI safety and governance framework is indispensable for maintaining the legitimacy, effectiveness, and sustainability of the governance system.**

### (b) Clarifying Roles and Responsibilities of Multiple Stakeholders in AI Governance

**Clearly defining roles and responsibilities serves as the cornerstone for fostering constructive interactions and long-term collaboration among multiple stakeholders.** Within the global AI safety and governance framework, different actors should, while respecting their differences, leverage their strengths and fulfill their respective responsibilities. Only through such efforts can an orderly and well-coordinated global AI safety and governance framework be achieved.

**National governments** should assume the primary responsibility for leading and regulating AI safety and governance, and establish comprehensive and effective national frameworks to address the sovereignty, safety, and ethical challenges posed by AI technologies. This in-cludes formulating and dynamically updating AI-related laws, regulations, and policy standards; instituting safety testing, evaluation, and certification systems for high-risk AI systems; establishing cross-departmental regulatory coordination mechanisms to ensure life-cycle risk monitoring, management, and control in AI research, development, and deployment; optimizing the layout of critical infrastructures such as computing power and data, and guiding orderly market participation and healthy industrial development; and, at the international level, actively participating in and promoting the development of a global AI safety and governance framework, and fostering fair, inclusive, and mutually beneficial cooperation mechanisms.

**International organizations** should play a key role in facilitating global dialogue, coordination, and rule alignment, and promote the formation of a more coherent international governance framework. This includes building global multilateral dialogue and negotiation platforms, and taking the lead in formulating international standards on AI safety, ethics, and interoperability; establishing cross-border review and oversight mechanisms to monitor and evaluate the implementation of national AI safety and governance policies and commitments; providing assistance and capacity-building support to countries with weaker technological capacity to bridge the global digital governance divide; and coordinating fragmented mechanisms to promote agenda complementarity and policy coherence among institutions in rule-making, risk assessment, and standard-setting.

**Enterprises** should assume corresponding responsibilities for ethics, safety, and governance in accordance with their positions within the industrial chain. For research and development enterprises, safety, transparency, and compliance requirements should be embedded in the design and development stages of frontier AI systems. For application enterprises, adaptation reviews and risk assessments should be conducted during the deployment and operation phases to ensure system robustness in areas such as public services and production safety, and to prevent misuse and abuse. All enterprises should fulfill their information disclosure obligations by proactively disclosing the technical characteristics and potential impacts of AI systems, accepting social supervision and compliance audits, and ensuring that technological innovation aligns with social responsibility.

**Research institutions** should undertake the critical functions of AI knowledge production and independent evaluation, providing objective scientific evidence and policy solutions for AI safety and governance. They serve to bridge the knowledge gap between technological frontiers and policy-making, thereby ensuring that decisions are scientifically sound and effective. Specifically, this involves independently conducting technical safety testing, socio-ethical impact assessments, and long-term risk forecasting of frontier AI systems; offering regulators science-based policy advice, references for standard-setting, and compliance assessment tools; exploring more ethical and safe technological pathways; and promoting interdisciplinary collaboration and consensus-building.

**Civil society and the public** are the value guardians and essential oversight forces in AI safety and governance. They should play a key role in value articulation and external supervision to ensure that governance processes remain transparent and inclusive. Specifically, civil society organizations can conduct independent research, promote ethical advocacy, and deliver public education to enhance risk awareness and promote value pluralism. The public can provide feedback, participate in social discussions, and monitor policy implementation to ensure that governance agendas respond to public interests and to prevent the exacerbation of social injustice or the erosion of public interests.

**At the institutional level, the role definition and responsibility boundaries of multiple stakeholders need to be further clarified through multi-level mechanisms. Regarding responsible entities,** differentiated institutional tools should be implemented: countries should participate in multilateral consultations and compliance reviews, and establish regular responsibility-reporting mechanisms; enterprises should fulfill mandatory transparency obligations, subject themselves to algorithmic safety audits and data protection compliance assessments; and research institutions and international organizations should rely on independent ethical review and technical evaluation to strengthen the standardization and compliance of research processes. **Regarding responsibility scope,** a multi-tiered and complementary rule coordination system should be developed. This governance architecture should be grounded in universally applicable global rules, supported by regional norms, and supplemented by industry standards. It should promote mutual recognition and alignment between international principles and domestic legislation, as well as between general requirements and context-specific regulations. Standard translation and adaptation mechanisms should be leveraged to reduce compliance costs and enhance governance effectiveness. **Regarding responsibility intensity,** institutional enforceability should be strengthened through a combination of "soft" and "hard" approaches: building voluntary value consensus and commitment through global ethical guidelines and industry conventions, while establishing binding compliance assurance systems through international treaties under the UN framework, regional agreements, and regulatory cooperation memoranda.

### (c) Building Effective Mechanisms for Multi-Stakeholder Coordination and Implementation

**Clarifying the roles and responsibilities of stakeholders serves as a foundational prerequisite, while an implementation mechanism that strengthens effective coordination is key to ensuring that global AI safety and governance are both enforceable and sustainable.** At present, the international community continues to face systemic challenges in areas such as compliance verification, risk monitoring, dispute resolution, and standard mutual recognition: **First, the lack of effective verification and oversight mechanisms** has made it difficult to supervise and implement commitments in areas such as AI safety, data governance, and algorithmic ethics. **Second, significant gaps remain in risk monitoring and information-sharing systems,** making it difficult to promptly identify and warn against systemic risks in model development, data usage, and computational resource allocation. **Third, dispute resolution mechanisms lack authority and timeliness,** and conflicts of interest among diverse actors lack efficient and authoritative mediation and arbitration channels. **Fourth, progress in cross-border and**

**cross-sectoral standards recognition has been slow,** constraining institutional alignment and execution efficiency in multilateral cooperation.

**Ensuring effective multi-stakeholder coordination and responsibility implementation in global AI safety and governance requires systematic and institutionalized mechanisms. Efforts should focus on the following four aspects: First,** establish strong and authoritative compliance review and oversight mechanisms. Regularly evaluate and publicly assess the commitments of countries and international organizations regarding rule adherence, policy implementation, and development assistance, thereby enhancing transparency and credibility in the implementation process. **Second,** establish a comprehensive and efficient risk monitoring and information-sharing mechanism. Require enterprises and research institutions to conduct standardized risk assessments and fulfill information-sharing obligations across key stages such as model development, computational resource allocation, and data usage, forming a dynamic, whole-chain early-warning network. **Third,** set up efficient and authoritative transnational dispute mediation and arbitration procedures. Provide binding channels for dispute resolution among governments, enterprises, research institutions, and the public, mitigating collaboration difficulties arising from unclear responsibilities or conflicting interests, and safeguarding the stability and effectiveness of AI safety and governance processes. **Fourth,** actively promote international standards recognition and regulatory alignment. Facilitate the development of compatible and mutually recognized normative frameworks in key areas such as safety evaluation, data governance, and ethical review, reducing institutional barriers and compliance costs, and enhancing the coordination and effectiveness of global AI safety and governance.

— PART —

# 05

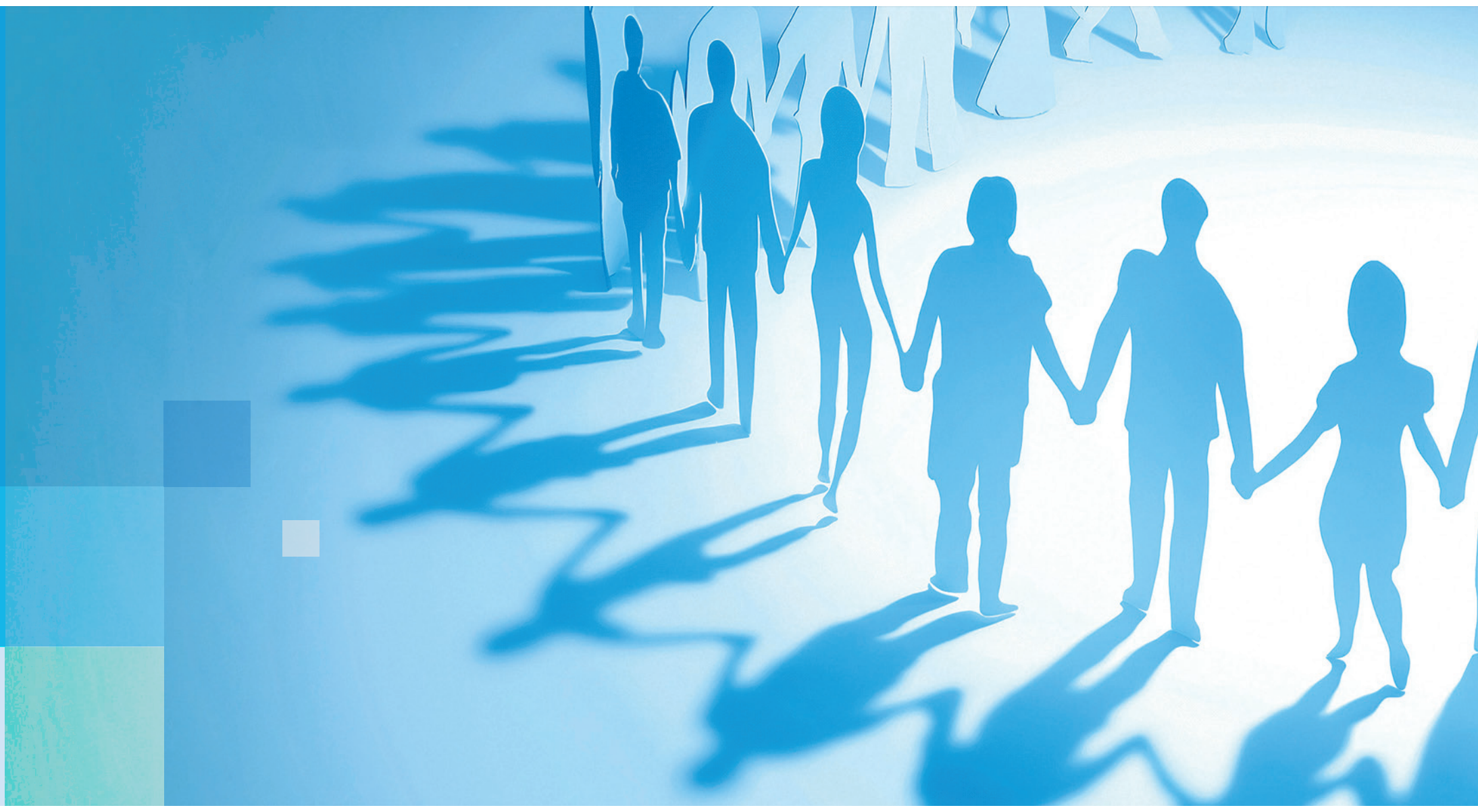

## Towards Our Future: Practicing Multilateralism and Building a Community with a Shared Future for Humankind

This chapter analyzes the impacts of of trends such as geopolitical competition and technological blockades on global cooperation, points out that a "People-Centered and AI for Good" approach serves as the fundamental value consensus for global AI governance. It emphasizes that with the United Nations at its core, efforts should focus on integrating existing governance resources, promoting cross-institutional coordination, and enhancing implementation capabilities to establish an authoritative and efficient global AI safety and governance framework.

### (a) Building and Implementing a Global Consensus towards the Common Good of Humanity

**Short-sighted actions driven by geopolitics are undermining the long-accumulated joint efforts of countries in AI governance and global cooperation.** In recent years, some countries have regarded AI technology as a key resource for strategic competition, adopting measures such as export controls, technological blockades, and supply chain decoupling, and prioritizing their own security and interests over global common goods. This policy orientation centered on prevention and confrontation is fragmenting the long-established transnational innovation networks and risk response systems. Meanwhile, the global AI governance framework is showing an increasingly prominent trend of fragmentation: the practice of formulating exclusive access standards and governance rules based on "small cliques" is on the rise. Bilateral agreements and regional bloc arrangements have weakened the coordination and openness of international rules, and eroded the breadth and inclusiveness of multilateral cooperation. Against this backdrop, the trust deficit between countries has further widened: sensitive issues such as algorithmic transparency, cross-border data flow, and computing resource sharing have been highly politicized,

information channels have been restricted, and as a result, transnational risk assessment and crisis response capabilities have been severely constrained.

**While countries differ in their institutional models and interests, the "People-Centered and AI for Good" approach remains the broad foundation for global consensus on AI safety and governance. First,** this consensus is highly endogenous: regardless of differences in technological development paths, safeguarding human dignity, promoting social well-being, and ensuring technological safety have always been the core demands for the policy legitimacy and social contract of most countries. Even against the backdrop of major-power strategic competition, technological innovation must ultimately serve the public interest and social progress—a shared goal widely recognized by all parties. **Second,** the cross-border and spreading nature of technological risks creates a natural resonance among countries on issues such as algorithmic transparency, privacy protection, and secure deployment. No single country can address global challenges alone, making cooperation the most rational choice. **Finally,** the long-term accumulation of multilateral frameworks and mechanisms provides institutional support for upholding this value consensus. Initiatives like the "Global Digital Compact" and UNESCO's "Recommendation on the Ethics of Artificial Intelligence" have exerted broad influence at the conceptual and normative levels, laying the groundwork for future alignment of rules, capacity building, and collaborative risk response.

**To implement the consensus of "People-Centered and AI for Good", it is essential to establish a working mechanism that balances scientific rigor with inclusiveness and promotes a pragmatic pathway of "Scientific Assessment + Tiered Dialogue."**

**Within the framework of the United Nations, it is recommended that the Independent International Scientific Panel on AI assume a guiding and coordinating role, while advancing the development of a global public service system for AI science communication accessible to all.** Building upon its mandate to provide professional advice to the UN system and global policymakers, the panel should further take on the responsibility of guiding international efforts in AI science communication. It should integrate relevant resources both within and beyond the UN system, leverage its professional authority and the multilateral advantages of the UN platform, and gradually form a broadly accessible, well-structured, and sustainable framework for public science communication on AI. This framework should adopt multilingual, multimedia, and cross-cultural communication strategies to support member states—particularly developing countries—and empower youth, educators, and community organizations to enhance their understanding of and participation in AI development and governance. By strengthening their rights to information, expression, and participation within the multi-stakeholder framework for global AI safety and governance, these efforts will help reinforce the social foundations and public trust essential to AI governance, thereby contributing to a more inclusive, equitable, and sustainable global governance framework.

**Meanwhile, it is necessary to gradually introduce the principle of risk-based, tiered, and classified dialogue into the existing United**

**Nations mechanisms for AI discussions, ensuring that the form and level of engagement are precisely aligned with the nature and magnitude of different risks.** With the support of the Independent International Scientific Panel on AI, a mechanism for frontier risk identification, classification, and early warning could be established, serving as the basis for a structured, multi-tiered dialogue framework. For regular or manageable risks, routine exchanges could be conducted at the expert and technical levels. For issues involving specific sectoral risks, relevant international and regional professional organizations should be fully engaged. For highly sensitive risks or those with potential systemic implications, the consultation format should be elevated as appropriate to a high-level policy dialogue mechanism involving national representatives and, when necessary, senior government leaders. This approach would help integrate major AI risk issues into national strategic agendas and scientific perspectives, promote policy coordination and resource integration, enhance public understanding, and strengthen cross-level responsiveness. Ultimately, it would reinforce the coherence, adaptability, and crisis response capacity of the global AI governance framework under the United Nations system.

### (b) Jointly Building a UN-Centered Global System for AI Safety and Governance

**As the most inclusive global governance platform, the United Nations plays an irreplaceable role in constructing the global AI safety and governance system.** As the most universal, authoritative, and representative intergovernmental organization in today's world, the UN, with the participation of nearly 200 member state governments, has an inherent advantage in coordinating diverse interests and balancing multiple concerns. Its legitimacy is founded on the shared principles and values enshrined in the "Charter of the United Nations". With the accelerated development of AI technology, the cross-border spillover of risks, and the growing fragmentation and institutional frictions in global governance, the UN's coordinating and leading role has become increasingly prominent. On the one hand, the UN system possesses abundant institutional resources. Bodies and initiatives such as the UN General Assembly, the UN Secretary-General's "Global Digital Compact" process, UNESCO, and the ITU provide multidimensional support for forging value consensus, unifying technical standards, strengthening risk governance, and advancing capacity-building cooperation. On the other hand, the UN's multilateral inclusiveness enables it to serve as a bridge between developed and developing countries, promoting the formation of a global rule-based framework that balances safety, development, and fairness.

**To address this coordination challenge, it is urgent to integrate and optimize AI governance resources within the UN system and continuously promote the building and deepening of consensus among all parties.** Currently, AI-related governance functions within the UN system are distributed across multiple institutions and thematic agendas, and their coordination mechanisms still have room for further strengthening. It is necessary to further strengthen central coordination and holistic integration to enhance synergistic effectiveness. The existing multiple mechanisms each have their own areas of focus: UNESCO promotes the formulation of AI ethical norms; The ITU is responsible for technical standards and infra-

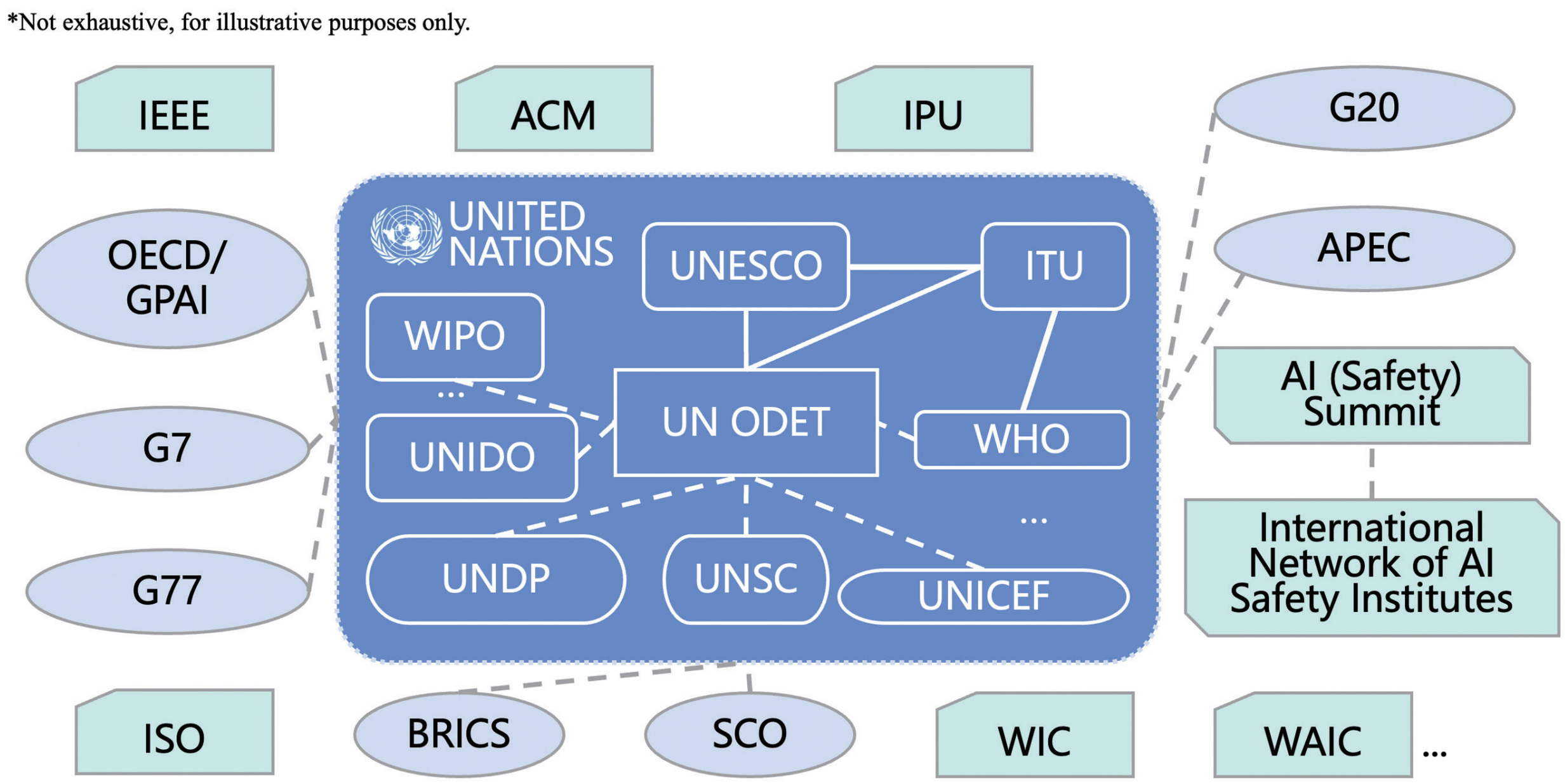

Figure 3: Schematic diagram of the relationship between UN systems and international organizations related to AI governance

structure interconnection; The Human Rights Council focuses on the protection of privacy and digital rights; and the Office of the UN Secretary-General's Envoy on Technology leads the "Global Digital Compact" process. These mechanisms have accumulated rich experience in their respective fields, but there are still deficiencies in overall coordination and linkage, preventing governance resources from functioning in full synergy. To this end, under the UN framework, it is necessary to establish an inter-institutional AI governance coordination mechanism to achieve agenda alignment and policy integration among different institutions. With the "Global Digital Compact" as the central platform, key issues such as value consensus, technical standards, risk governance, and capacity building should be systematically incorporated into a unified agenda.

**Looking ahead, the international community could further explore, under the framework of the United Nations, the establishment of a global AI safety and governance platform with stronger coordination and implementation capacity.** The platform could operate with a mandate from the UN General Assembly to ensure its legitimacy and representativeness, and, based on the UN Office for Digital and Emerging Technologies (ODET) and other relevant UN bodies, establish an inter-institutional agenda coordination system[115]. At the same time, it could draw on the successful experiences of other global governance mechanisms—such as the Intergovernmental Panel on Climate Change (IPCC) in the field of climate change, the normative system of the International Civil Aviation Organization (ICAO), the peer review mechanism of the Convention on Nuclear Safe-

115. The Intergovernmental Panel on Climate Change. Source: https://www.ipcc.ch/

ty, and the emergency response framework of the International Health Regulations (IHR)—to promote cross-system integration across multiple dimensions, including technical standards[116], risk governance[117], ethical review [118], and capacity building[119]. The platform should possess functions of coordination, deliberation, standard-setting, and necessary implementation and oversight authority, and could gradually establish the compliance obligations of member states through instruments such as international conventions[120]. It should also strengthen cross-border risk information-sharing[121] and emergency response mechanisms[122], and promote the international mutual recognition and implementation of technical standards [123]. Through these efforts, the platform would enhance the overall coordination and execution capacity of global AI governance, enabling it to serve as a core force for coordinating global safety and governance resources, balancing diverse interests, and promoting effective implementation.

**At the intersection of an era marked by rapid technological evolution and profound transformation of the global order, AI safety and governance are vital to the shared future for humankind.** With the United Nations at its core, the international community should continue to build and deepen consensus, and mobilize resources in accordance with the principles of sovereign equality, international rule of law, and multilateralism. It should uphold a people-centered, inclusive, cooperative, and results-oriented value orientation, and work together to forge a new global governance framework that balances safety, development, equity, and inclusiveness. Amid an era of both opportunities and challenges, all parties should jointly open an intelligent future that is inclusive, shared, and sustainable.

116. “Convention on International Civil Aviation (Chicago Convention)”, which establishes the fundamental legal framework and principles for international civil aviation activities and created the International Civil Aviation Organization (ICAO). Source: https://www.icao.int/convention-international-civil-aviation-doc-7300
117. “Convention on Nuclear Safety”, which ensures the safe operation of civilian nuclear power plants and promotes continuous improvement of nuclear safety through peer reviews. Source: https://www.iaea.org/topics/nuclear-safety-conventions/convention-nuclear-safety
118. “WHO Framework Convention on Tobacco Control”. The first global public health treaty negotiated under the auspices of the World Health Organization, which aims to address the global tobacco epidemic by reducing tobacco consumption and supply. Source: https://wkc.who.int/resources/publications/i/item/9241591013
119. United Nations, “Enhancing international cooperation on capacity-building of artificial intelligence”. Source:https://docs.un.org/en/A/RES/78/311
120. “United Nations Framework Convention on Climate Change (UNFCCC)”, the foundational legal framework for global climate change response. Source: https://www.un.org/climatesecuritymechanism/en/united-nations-framework-convention-climate-change-unfccc-and-climate-peace-and-security
121. “Convention on Early Notification of a Nuclear Accident”, which requires State Parties to provide immediate notification in the event of a nuclear accident that may have radiological safety implications for other countries. Source: https://www.iaea.org/topics/nuclear-safety-conventions/convention-early-notification-nuclear-accident
122. WHO, “The International Health Regulations (IHR)”—a legally binding global framework for public health security designed to help countries prevent and respond to public health emergencies of international concern. Source: https://www.who.int/health-topics/international-health-regulations#tab=tab_1
123. “International Convention for the Safety of Life at Sea (SOLAS)”, an international treaty under the auspices of the International Maritime Organization (IMO) that aims to establish uniform standards for the construction, equipment, and operation of merchant ships to ensure safety. Source: https://www.imo.org/en/knowledgecentre/conferencesmeetings/pages/solas.aspx

## Appendix: Proposed Recommendations for a Global AI Safety and Governance Framework

| Dimension | Key Issues | Proposed Mechanisms | Expected Outcomes | Relevant International Experience |
|---|---|---|---|---|
| **Responding to the Rapid Transformation and Major Risks of AI** | Uncertain directions of technological breakthroughs, insufficient risk foresight, and lack of relevant indicators and capabilities | **Technology tracking and risk early warning collaboration:** establish cross-border technology monitoring networks, promote information sharing, and conduct regular, dynamic assessments of technological progress | Accurately capture breakthrough directions, identify potential risks in advance, and build a global front line for AI safety | IPCC's interdisciplinary data-sharing platform and predictive mechanisms; IAEA's verification and reporting mechanisms (early risk detection) |
| | Slow rule-making and updating processes; regulatory measures differ across countries and regions | **Dynamic updating and mutual recognition mechanism for governance rules:** institutionalize periodic reviews and updates of AI safety standards to ensure rule interoperability and dynamic adaptation | Eliminate regulatory gaps caused by divergent national standards and strengthen the foundation for global governance collaboration | IPCC's model of linking scientific assessments with policy recommendations (science-informed, dynamic policy updates) |
| | Lack of systematic risk assessment and intervention tools; immature model safety monitoring; limited regulatory effectiveness | **Common safety evaluation tools and platform ecosystem:** jointly develop safety testing platforms, benchmark datasets, and risk assessment tool libraries to build a globally shared technical toolbox | Address weak regulatory capabilities by providing unified technical support for national regulators | WHO's International Health Regulations (IHR) epidemic information reporting and cross-regional tracking mechanisms (multi-source data + rapid feedback) |
| | Insufficient consensus on risk perception; inconsistent standards and regulatory approaches | **International consensual risk governance framework:** coordinate national risk classification standards through multilateral consultation | Establish an internationally recognized risk categorization system to enhance cross-border risk identification | IAEA's International Nuclear and Radiological Event Scale (INES) |
| | Underdeveloped cross-border risk prevention and traceability mechanisms | **Cross-border content governance collaboration:** leverage existing international standard organizations to promote interoperability among content authentication technologies | Build a global AI-generated content traceability and authentication network with shared databases | ISO/IEC's international standard system promoting content authentication and mutual recognition |
| | Difficulties in cross-border law enforcement cooperation; lack of standardized cooperation procedures | **Cross-border law enforcement cooperation mechanism:** standardize information sharing, joint investigation, and evidence collection | Build monitoring and joint-response mechanisms to counter AI misuse and enhance the ability to respond to cross-border violations | Budapest Convention on Cybercrime — cross-border enforcement and evidence cooperation mechanisms |
| | Limited participation in international joint assessments of frontier risks; standards and procedures confined to a few countries | **International cooperation and multilateral dialogue mechanisms:** foster consensus on addressing potential AI runaway risks | Prevent technological misuse and loss of human control, ensuring AI development remains safe, reliable, and under human oversight | Biological Weapons Convention and Chemical Weapons Convention — multilateral cooperation and transparency review mechanisms |
| | Lack of transparency in frontier models; poor explainability; difficult to detect misuse in time | **Frontier model monitoring and evaluation:** assess the capabilities and misuse risks of high-risk models and establish cross-border reporting systems | Enable timely identification and warning of potential threats, enhancing transparency and cross-border trust | IAEA's verification and reporting mechanisms; Convention on Early Notification of a Nuclear Accident (urgent notification duties) |
| | Unequal distribution and potential abuse or monopolization of key resources (compute power, data, algorithms, chip manufacturing) | **Critical resources governance:** develop international consensus on coordinated management of critical AI resources such as computing power, data, algorithms and chips | Reduce risks of misuse from the source and promote equitable and controllable access to resources | WTO's coordination mechanisms (anti-monopoly and fair resource access); ICAO and IMO standardization frameworks (cross-system interoperability and accountability tracking) |
| | Lack of emergency response capacity for extreme risk scenarios; individual countries unable to manage transnational crises | **Emergency intervention and joint response mechanisms:** establish rapid multinational coordination protocols (e.g., suspending risky experiments, halting model deployment, freezing financial flows) | Ensure rapid and effective responses in high-risk situations to maintain control over critical systems and prevent global catastrophic outcomes | WHO's IHR Emergency Committee mechanism (rapid emergency response) |
| **Balancing Global Development and Governance Demands in AI** | Difficulty building international consensus; divergent positions on key governance issues; high coordination costs | **Open and inclusive multilateral consultation platforms:** ensure diverse participation in agenda setting, rule-making, and standard-setting processes | Foster global consensus and inclusive participation to enhance legitimacy and fairness in global governance | United Nations Framework Convention on Climate Change (UNFCCC): equal participation, common but differentiated responsibilities; WTO multilateral negotiation system |
| | Challenges in advancing unified regulatory frameworks; lack of cross-border mutual recognition and coordination mechanisms | **Practical and effective international cooperation:** promote capacity building, knowledge sharing, and resource support for developing countries | Bridge the global North-South divide, promote technology inclusivity, improve infrastructure, and develop human capital | WHO's capacity-building assistance mechanism; WTO's Trade-Related Technical Assistance (TRTA) activities |
| | Imbalance in governance participation; dominance by technologically advanced countries | **Fair and transparent decision-making procedures:** establish accountable and transparent coordination mechanisms | Reduce information asymmetry, enhance mutual trust, and build institutional foundations for sustained global risk governance | WTO's Dispute Settlement Mechanism (clear rules, transparent processes) |
| | Large gaps in technological capability; developing economies lack structural advantages in core resource areas | **Technology sharing and transfer:** promote controlled cross-border flows of compute, algorithms, data, and applications | Enable developing countries to leapfrog in technological capacity, narrowing the global digital divide | WTO's TRIPS Agreement provisions on technology transfer |
| | Insufficient financial investment in developing countries; limited public resources and self-sufficiency in technology | **Innovative financing models:** establish AI development funds through UN, World Bank, and regional development banks | Provide sustained funding for R&D, infrastructure, and public data, mitigating structural financial deficiencies | Global Environment Facility (GEF); World Bank's digitalization programs |
| | Short-term capacity-building support unsustainable for least-developed regions | **Long-term capacity-building cooperation mechanisms:** provide large-scale assistance in talent cultivation, regulatory frameworks, standard alignment, and risk governance | Strengthen local governance and adaptive development capacity, improving global AI safety resilience | WHO's capacity-building assistance mechanism; WTO's Trade-Related Technical Assistance (TRTA) activities |
| | Lack of credit and enforcement mechanisms at the international level; weak compliance and low cost of violations | **Dynamic governance participation evaluation:** continuously assess national participation and compliance in global AI governance, linking violations to international reputation | Raise the cost of non-compliance and enhance rule enforcement | WTO's Trade Policy Review Mechanism; IAEA's compliance supervision mechanism |
| | Developed countries monopolize discourse power, increasing compliance costs for latecomers | **Decision-making hearing mechanism:** uphold representativeness, transparency, procedural justice, and consensus-based decision-making | Strengthen the legitimacy, inclusiveness, and trust of governance decisions | UN Universal Periodic Review (UPR); WTO's Trade Policy Review Mechanism |
| | Developing countries bear high compliance burdens without real influence in decision-making | **Differentiated responsibility-sharing mechanism:** assign obligations based on development levels, technological capacity, and governance foundation | Achieve a balance between inclusiveness and effectiveness; promote equitable responsibility distribution | UNFCCC's Common but Differentiated Responsibilities principle; WTO's Special and Differential Treatment provisions |
| **Promoting Coordinated and Effective Multi-Stakeholder Action** | Weak implementation and accountability by nations and international organizations | **Compliance review and oversight mechanisms:** track rule-making, policy implementation, and capacity-building commitments | Enhance transparency, accountability, and credibility of governance implementation | IAEA's verification and compliance mechanisms; WTO's Trade Policy Review Mechanism |
| | Difficulty in identifying and warning of risks during R&D and deployment | **Risk monitoring and information sharing mechanisms:** require enterprises and research institutions to conduct standardized risk assessments and share relevant information | Build a multi-layered, end-to-end dynamic risk warning network | WHO's IHR epidemic reporting system; ICAO security occurrences and incidents reporting systems |
| | Conflicting stakeholder interests; blurred responsibilities causing governance gridlock | **Dispute resolution and arbitration procedures:** provide efficient, impartial, and enforceable dispute channels for governments, enterprises, researchers, and the public | Avoid governance deadlocks and maintain coordination stability and effectiveness | WTO's Dispute Settlement Mechanism; York–Antwerp Rules (YAR) on liability allocation |
| | Slow progress in mutual recognition of cross-border and cross-sectoral standards | **Standards recognition and regulatory alignment mechanisms:** promote interoperability in algorithm safety, data governance, and technical evaluationm, and form a compatible normative framework | Reduce institutional friction and duplication, improving global coordination and enforcement efficiency | ICAO and IMO international standard systems; ISO's Mutual Recognition Agreement or Arrangement (MRA) |
| **Practicing Multilateralism and Building a Community with a Shared Future for Humankind** | Low public scientific literacy and limited participation in governance; policymakers' scientific literacy needs improvement | **Science communication system:** led by the UN Independent International Scientific Panel on AI, develop a multilingual, cross-media global communication network targeting youth, educators, and grassroots organizations—especially in developing countries | Strengthen democratic foundations and public trust in AI governance; enhance global awareness and participation | UNESCO's global education and communication programs |
| | Single-layer risk dialogue system; insufficient response efficiency | **Tiered and integrated dialogue mechanism:** establish regular expert-level exchanges for routine risks, and high-level consultations (involving state representatives or heads of government) for critical risks to elevate them to strategic agendas | Improve major risk response efficiency and reinforce UN multilateral governance mechanisms | IPCC's tiered scientific assessment and reporting system |

## World Internet Conference Specialized Committee on Artificial Intelligence

The Specialized Committee on Artificial Intelligence, as the first professional branch of World Interne Conference, was established in 2024, including three programs: Standards Program, AI Safety and Governance Program, and Industry Program. The Committee brings together leading experts and professionals in the field of AI from international organizations, think tanks, research institutes, professional associations and industry, which is dedicated to fostering international cooperation and coordinated development, and aims to facilitate global sharing of AI achievements. Through thematic seminars, outcome sharing, and initiative releases, it continuously consolidates international consensus and supports inclusive and sustainable AI development.

## Advancing a Global Framework for AI Safety and Governance for the Well-being of Humanity

Follow us on Facebook: @wicinternet

Follow us on X: @wicinternet

世界互联网大会
World Internet
Conference